%% file: ms.tex
\documentclass[usenatbib]{mn2e}
\usepackage{natbibmnfix, graphicx, times, amsmath, epsfig, amssymb}
\include{defns}

\pdfoutput=1

\begin{document}

\title[The clustering of LAEs during reionization]{The clustering of Lyman alpha emitters at $\pmb{z\approx7}$: \\implications for reionization and host halo masses}
\author[Sobacchi \& Mesinger]{Emanuele Sobacchi$^1$\thanks{email: emanuele.sobacchi@sns.it} \& Andrei Mesinger$^1$  \\
$^1$Scuola Normale Superiore, Piazza dei Cavalieri 7, 56126 Pisa, Italy\\
}

\voffset-.6in

\maketitle

\begin{abstract}
The \lya\ line of high-redshift galaxies has emerged as a powerful probe of both early galaxy evolution and the epoch of reionization (EoR). Motivated by the upcoming wide-field survey with the Subaru Hyper Suprime-Cam (HSC), we study the angular correlation function (ACF) of narrow-band selected, $z\approx7$ \lya\ emitting galaxies (LAEs).  The clustering of LAEs is determined by both: (i) their typical host halo masses, $\bar{M}_{\rm h}$; and (ii) the absorption due to a patchy EoR, characterized by an average neutral fraction of the intergalactic medium, $\avenf$.
We bracket the allowed LAE ACFs by exploring extreme scenarios for both the intrinsic \lya\ emission and the large-scale pattern (i.e. morphology) of cosmic ionized patches in physical EoR models.
Independent of the EoR morphology, current $z\approx7$ ACF measurements constrain $\avenf \lsim 0.5$ (1-$\sigma$).   
We also find that the low values of the currently-observed ACF imply that LAEs are hosted by relatively small dark matter halos: $\bar{M}_{\rm h} \lsim 10^{10} \Msun$, with corresponding duty cycles of $\lsim$ few per cent. These values are over an order of magnitude lower than the analogous ones for color-selected, Lyman break galaxies, suggesting that
$z\approx7$ narrow-band LAEs searches are preferentially selecting young, star-burst galaxies, residing in less massive halos. The upcoming Ultra Deep campaign with the HSC will significantly improve constraints on both the EoR and LAE host halos.
\end{abstract}


\section{Introduction}
\label{sec:intro}

Although it is the last major phase change in the history of our Universe, the epoch of reionization (EoR) remains poorly explored.  The EoR is expected to be fairly extended and patchy, with the growth of cosmic HII regions sourced by the birth of an increasing number of early galaxies.  Understanding this complex process requires accurate statistics.  Luckily, efforts are underway to increase the sample of high-redshift ($z\gsim6$) objects serving as EoR probes, such as quasars and galaxies, hopefully resulting in an unambiguous EoR detection.

Some of the most important of these efforts are focused on Lyman alpha emitters (LAEs).  Due to the significant \lya\ damping wing opacity of the neutral intergalactic medium (IGM), Lyman alpha emission from galaxies located close to the edges of cosmic HII regions will be strongly attenuated.  Conversely, 
since their Ly$\alpha$ photons can redshift even further away from resonance before reaching the neutral IGM,
galaxies far from these edges will be less affected and therefore easier to observe. Thus the EoR will impact both the observed number of LAEs as well as their clustering properties.

The main difficulty in using LAEs as an EoR diagnostic lies in the fact that we do not understand their intrinsic properties, such as their \lya\ luminosity, the profile of the \lya\ lines, and the absorption within the local circumgalactic medium (CGM).  These intrinsic properties can be degenerate with an EoR signature.
Lacking this a priori information on intrinsic properties, most modern studies focus on redshift evolution, with the implicit assumption that most of the evolution is due to the EoR: a change in the volume-average neutral fraction of the IGM, $\avenf$.  Indeed, both the fraction of drop-out galaxies with strong \lya\ emission and narrow-band LAE surveys show a drop in these populations at $z\gsim6$, countering empirical trends from lower redshifts (e.g. \citealt{Ouchi10, Fontana10, Stark10, Kashikawa11, Caruana12, Treu13, Konno14, Schenker14, Caruana14, Pentericci14, Cassata15}).  However, small-number statistics, as well as ignorance of the evolution in intrinsic \lya\ emission and its relation to continuum selected galaxies prevents a robust claim of an EoR detection (e.g. \citealt{DW10, DMF11, BH13, Jensen13, Dijkstra14, TL14, Mesinger15}).

An alternative EoR diagnostic is the observed clustering of LAEs. Clustering could be a more powerful EoR probe than pure number counts, for several reasons. Firstly, the intrinsic (not modified by the IGM) clustering of LAEs is determined almost entirely by their host halos.  The correlation functions of dark matter (DM) halos are well understood and only evolve by factors of $\sim$few over the range of possible host halo masses;  this is in contrast to uncertainties in baryonic physics (e.g. the \lya\ escape fraction, star-formation rates) which can have an order unity impact on the observed number density.  Secondly, the galactic environment (including self-shielded systems and accretion flows), while possibly having a large (degenerate with that of the EoR) imprint in the number density and its evolution (e.g. \citealt{DW10, Finkelstein12, BH13, Mesinger15}), will have a much weaker impact on the the observed clustering signature on the large ($\sim$tens Mpc) scales relevant for the EoR (e.g. \citealt{Crociani11, WD11, Mesinger15}).  

The above means that the uncertainties associated with the intrinsic clustering signature of LAEs are much smaller than those of their intrinsic number densities.  This could allow us to constrain the EoR properties from a just a single, high-$z$ clustering measurement (e.g. \citealt{FZH06, McQuinn07LAE}).  Indeed, using a counts-in-cell statistic, \citet{MF08LAE} showed that the LAE clustering induced by patchy reionization is so strong during the first half of the EoR, that it exceeds even the worst case scenario of LAEs being hosted by the most massive, most {\it intrinsically}-clustered halos.  In other words, the EoR-induced clustering would have an {\it unambiguous imprint} during the first stages of reionization, which {\it cannot be replicated} by even ad-hoc models for relating intrinsic \lya\ luminosities to their host halos $L_{\alpha}^{\rm intr} \leftrightarrow M_{\rm h}$.  Such a measurement would provide the much-needed ``smoking-gun'' proof for reionization.

Unfortunately, we are unlikely to have a measurement of the \lya\ clustering at the redshifts high enough to probe the first half of reionization (which best-fit \citealt{Planck15} models place at $z\gsim9$, albeit with large error bars).  Indeed, preliminary estimates suggest that the Ultra Deep campaign with the HSC will only capture tens of LAEs at $z\approx7.3$ (compared to $\gsim1000$ at $z\approx6.6$; M. Ouchi, private communication). Hence, constraints using LAE clustering during the second half of reionization (as expected to be the case at $z\approx6.6$) must explore the degeneracy between which halos host LAEs (i.e. intrinsic clustering) and the patchy reionization-induced clustering.

In this study we compute the angular correlation functions (ACFs) of LAEs at $z\approx7$, using several EoR models as well as prescriptions for assigning intrinsic \lya\ luminosities to host halos ($L_{\alpha}^{\rm intr} \leftrightarrow M_{\rm h}$). Our work is similar to the recent predictions by \citet{Jensen14}; however, given the large modeling uncertainties at high redshifts, here we adopt a more systematic approach by using extreme scenarios both for EoR morphologies as well as the $L_{\alpha}^{\rm intr} \leftrightarrow M_{\rm h}$ relation in order to bracket the allowed parameter space.  Moreover, our framework uses the latest EoR simulations from \citet{SM14}, which include state-of-the-art sub-grid models of important physical processes like inhomogeneous recombinations inside small-scale structures.
As this work was nearing completion, a similar effort was also made by \citet{HDM15}.  
Our work also differs from this analysis, which derives EoR constraints using one specific LAE model and one specific reionization scenario driven by $z=6.6$ galaxies.

This study is motivated by the upcoming $z=6.6$ Subaru survey with the Hyper Suprime-Cam (HSC).  Current Subaru clustering measurements are suggestive of a mostly ionized universe at $z\approx7$ \citep{McQuinn07LAE, Ouchi10, Jensen13}. The upcoming HSC survey will result in up to a dramatic 30-fold increase in survey volume ($\sim$30 deg$^2$ for the deep survey; M. Ouchi, private communication), finally allowing us to statistically sample the EoR morphology \citep{TL14, Jensen14}.

This paper is organized as follows: in Section \ref{sec:model} we present our models for reionization and LAEs; in Section \ref{sec:results} we see how to recover robust limits on  $\bar{x}_{\rm HI}$ from clustering measurements; in Section \ref{sec:concl} we present our conclusions.
Throughout we assume a flat $\Lambda$CDM cosmology with parameters ($\Omega_{\rm m}$, $\Omega_{\rm \Lambda}$, $\Omega_{\rm b}$, $h$, $\sigma_{\rm 8}$, $n$) = ($0.28$, $0.72$, $0.046$, $0.70$, $0.82$, $0.96$), as measured by the Wilkinson Microwave Anisotropy Probe (WMAP; \citealt{Hinshaw13}), consistent with recent results from the Planck satellite \citep{Planck15}. Unless stated otherwise, we quote all quantities in comoving units.

\section{Model}
\label{sec:model}

The observed \lya\ luminosity of a given galaxy, residing in a DM halo of total mass, $\Mhalo$, can be written as:
\begin{equation}
L_\alpha = L_{\alpha}^{\rm intr} e^{-\tau_{\rm IGM}} > L_\alpha^{\rm min} ~ ,
\end{equation}
where $L_{\alpha}^{\rm intr}(\Mhalo)$ is the intrinsic (before passing through the IGM) \lya\ luminosity, $\tau_{\rm IGM}(\avenf)$ is the IGM optical depth along a given sightline, and $L_\alpha^{\rm min}(z)$ is the sensitivity threshold of the observation (e.g. $L_\alpha^{\rm min}=2.5\times10^{42}$ erg s$^{-1}$ for the upcoming HSC Ultra Deep survey at $z=6.6$).
Thus, modeling the clustering of observed LAEs requires two components: (i) a scheme for assigning intrinsic \lya\ emission to host DM halos, $L_{\alpha}^{\rm intr}(\Mhalo)$; and (ii) large-scale reionization simulations for determining the IGM opacity during the EoR, $\tau_{\rm IGM}(\avenf)$. 
Given that both of these are poorly-constrained, we adopt a systematic approach and explore extreme models for both (i) and (ii).  Below we describe each in turn.

\subsection{The intrinsic emission of LAEs}

The intrinsic \lya\ emission of high-$z$ galaxies depends on complicated physical processes inside the interstellar medium (ISM) and circumgalactic medium (CGM), and cannot be computed from first principles.  Instead, intrinsic luminosities can be predicted empirically using observations of the \lya\ equivalent widths at lower redshifts (where the LAE samples are sufficiently large), which relate the \lya\ luminosity to the UV continuum magnitude through some probability distribution function (see, e.g. \citealt{DW12} and references therein).  The UV continuum magnitude can then be related to the DM halo mass using standard abundance matching, and extrapolating down the faint end slope, under the assumption of a mass-independent duty cycle, $f_{\rm duty}$\footnote{Here we define a duty cycle as the fraction of halos of a given mass which host detectable LAEs, i.e. with L$_\alpha$ $\ge 2.5\times 10^{42}\text{ erg s$^{-1}$}$ for our fiducial mock survey.}
(e.g. \citealt{KF-G12}).  This procedure however is highly uncertain, relying on empirical, poorly-constrained extrapolations in both redshift and magnitude, as well as implicit assumptions about the mass and redshift scalings of distributions and duty cycles.

Here we adopt a different approach. We explore three parametric prescriptions relating the \lya\ luminosities to halo masses.  Although these models match current observational constraints, they are not intended to be overly realistic.  Rather they are intended to bracket the allowed range of the clustering signal.  Consistent with $z\sim4$ observations (e.g. \citealt{Gronke15}), we assume the \lya\ luminosity increases with halo mass:
\begin{equation}
  L_{\alpha}^{\rm intr}=L_\alpha^{\rm min}\left(\frac{M_{\rm h}}{M_{\rm \alpha}^{\rm min}}\right)^\beta\chi ~ ,
  \label{eq:LtoM}
\end{equation}
where $\chi$ is a random variable ($\chi=1$ with probability $f_{\rm duty}$ and $\chi=0$ otherwise), which accounts for the expected bursty star formation inside high-$z$ DM halos (e.g. \citealt{Lee09, WLO14}).  The normalization of the intrinsic luminosity in eq. (\ref{eq:LtoM}) is governed by the halo mass corresponding to the detection threshold, $M_{\rm \alpha}^{\rm min}$, and the shape of the LF can be parametrized with the power law index, $\beta$.  Our fiducial models assume $\beta=1$, though we also compute correlation functions with $\beta=2/3$ without notable changes to our conclusions.

Our systematic approach therefore makes use of three ``tuning knobs'' for the intrinsic luminosity of LAEs: (i) the \lya\ duty cycle, $f_{\rm duty}$, which shifts the LAE LFs up and down; (ii) the normalization, $M_{\rm \alpha}^{\rm min}$, which shifts the LAE LFs left and right, and (iii) the power-law scaling with mass, $\beta$, which flattens or steepens the LAE LFs.  For a given value of (ii) and (iii), changing (i) does not impact the LAE clustering properties, only the observed number density of LAEs. In our analysis of the $z=6.6$ clustering, we vary $f_{\rm duty}$ with the neutral fraction of the IGM during the EoR, in order to {\it fix the observed number density of LAEs to the already-constrained value} of $\nlea(z=6.6)$ = $4.1^{+0.9}_{-0.8}\times10^{-4}$ Mpc$^{-3}$ (for L$_{\alpha}^{\rm intr}$ $\sim 2.5\times 10^{42}\text{ erg s$^{-1}$}$; \citealt{Ouchi10}).  Therefore, a given value of $M_{\rm \alpha}^{\rm min}$ requires a higher value of $f_{\rm duty}$ earlier in reionization, in order to compensate for the added attenuation of the IGM.\footnote{The exception to this procedure is the ``most massive halo'' model below, which assumes a duty cycle of unity.  As the duty cycle cannot be further increased, for this model we allow lower observed number densities during the EoR.  However, as we shall see below, this extreme model is already ruled out.}

Instead, (ii) and (iii) do impact the intrinsic LAE clustering.  Their combined impact can be better parametrized in term of the
{\it luminosity-weighted} average host halo mass:
\begin{equation}
\bar{M}_{\rm h}=\left( \frac{\int_{M_{\rm \alpha}^{\rm min}}^{+\infty}M_{\rm h}^\beta n(M_{\rm h}) dM_{\rm h}} {\int_{M_{\rm \alpha}^{\rm min}}^{+\infty} n(M_{\rm h}) dM_{\rm h}} \right)^{-\beta}\;.
\end{equation}
where $n(M_{\rm h})$ is the halo mass function (HMF) at $z=6.6$ (e.g. \citealt{Jenkins01}).  The HMF is quite steep over the relevant range, so that $\bar{M}_{\rm h}$ is at most a factor of few higher than $M_{\rm \alpha}^{\rm min}$ for reasonable values of $\beta \sim 1$, with the faintest, most abundant LAEs dominating the intrinsic clustering.  Below, we keep $\beta=1$ and vary the normalization in order to change the intrinsic clustering properties, as parametrized by the luminosity-weighted mean halo mass.  Although they do not impact our clustering conclusions, we do show some models with different values of $\beta$, illustrating that more accurate LFs could constrain this scaling.

We take three models for $\bar{M}_{\rm h}$, spanning the allowed range:
\begin{itemize}
\item {\bf Most massive halos, $\bar{M}_{\rm h} \approx 2 \times 10^{11} \Msun$}: this relation maximizes the intrinsic clustering of LAEs by assuming that the observed LAEs are hosted in the most massive, most intrinsically clustered DM halos.  This is achieved by using a duty cycle of unity, $f_{\rm duty}=1$, which results in an average halo mass of $\bar{M}_{\rm h} \approx 1.8 \times 10^{11} \Msun$, when normalized to the observed LAE number density.

\item {\bf Least massive halos, $\bar{M}_{\rm h} \approx 3 \times 10^{9} \Msun$}: this relation is chosen to minimize the intrinsic clustering of LAEs.  LAE populate DM halos down to masses of $M_{\rm \alpha}^{\rm min}=10^{9} \Msun$.  This is an extreme value, corresponding to the conversion of all halo baryons into stars inside a single dynamical timescale, with all \lya\ photons escaping the galaxy\footnote{The minimum mass of the halos hosting visible LAEs (with L$_{\alpha}^{\rm intr}$ $\sim 2.5\times 10^{42}\text{ erg s$^{-1}$}$ can be crudely estimated by the following argument (c.f. \citealt{MF08LAE}): assuming that about 2/3 of the ionizing photons absorbed within the galaxy are converted into Ly$\alpha$ photons \citep{Osterbrock89}, one can write the conversion as L$_{\alpha}^{\rm intr}=0.67\times h\nu_{\alpha} \left(1-f_{\rm esc}\right)\dot{\rho}_\ast\varepsilon_\gamma T_{\rm \gamma,res}$, where $\nu_\alpha$ is the rest-frame Ly$\alpha$ frequency, fesc is the escape fraction of ionizing photons, $\dot{\rho}_\ast$ is the star formation rate (SFR), $\varepsilon_\gamma$ is the ionizing photon efficiency per stellar mass and $T_{\rm \gamma, res}$ is the fraction of Ly$\alpha$ photons which escape from the galaxy without getting resonantly absorbed. Assuming that galaxies convert a fraction, $f_\ast$, of their gas into stars over some mean time-scale, $t_\ast$, and that $f_{\rm esc}\ll 1$, one can write the above relation as:  L$_{\alpha}^{\rm intr} \sim 2 \times 10^{-12}\left(\frac{\varepsilon_\gamma f_\ast T_{\rm \gamma, res}}{t_\ast}\right)M_{\rm h}\text{ erg s}^{-1}$.  We obtain our most extreme model for L$_{\alpha}^{\rm intr}(M_{\rm h})$  by   maximizing $\varepsilon_\gamma f_\ast T_{\rm \gamma, res}/t_\ast$: $\varepsilon_\gamma = 6\times 10^{60}$ ionizing photons $M_\odot^{-1}$ ( corresponding to a PopII IMF with $Z=0.04\times Z_\ast$ from \citealt{Schaerer02}),  $f_\ast T_{\rm \gamma, res}=1$, $t_\ast\approx 200$ Myr corresponding to the dynamical time at $z\sim7$.
  These extreme assumptions result in the relation L$_{\alpha}^{\rm intr} \sim 2\times 10^{42}\left(M_{\rm h}/10^9 M_\odot\right)$ erg s$^{-1}$, corresponding to a minimum mass-scale $\sim 10^9M_\odot$ for the faintest LAEs in the Subaru Ultra Deep surveys.
  Nevertheless, this estimate is very rough, and it is theoretically possible to detect extremely young star-bursts in any halo massive enough to host a galaxy (above the atomic cooling threshold).  We explore this even-more-extreme possibility in the Appendix (Fig. \ref{fig:LAE_main_lowM}), showing that  our conclusions remain unchanged, since ACFs are very weak functions of halo mass this far below the knee of the HMFs.}
Matching the observed LAE number densities in this model requires a duty cycle $f_{\rm duty}\approx 0.001$, assuming a mostly-ionized Universe.

\item {\bf Moderately massive halos, $\bar{M}_{\rm h} \approx 2 \times 10^{10} \Msun$}: this relation is chosen as an intermediate between the two extremes above, and is in approximate agreement with the previous best-fit values obtained from the clustering of LAEs at $z\sim$6--7 (e.g. \citealt{Ouchi10}).  The corresponding duty cycle is $f_{\rm duty}\approx 0.02$, assuming a mostly-ionized Universe.
  
\end{itemize}

\begin{figure}
\vspace{+0\baselineskip}
{
\includegraphics[width=0.45\textwidth]{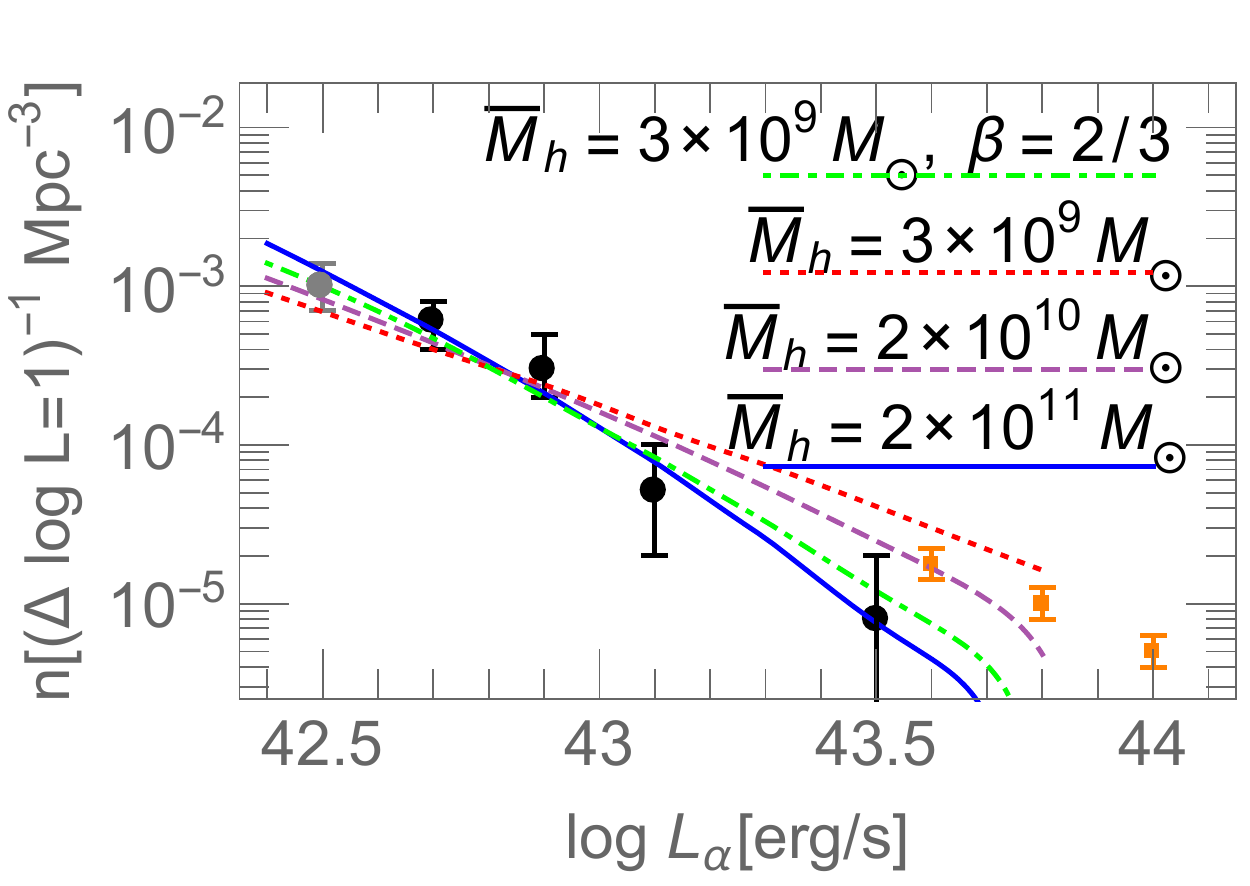}
}
\caption{LAE luminosity functions at redshift $z=6.6$; dots/squares represent observations by \citet{Ouchi10}/\citet{Matthee15}.  
  With the solid, dashed and dotted lines we show the simulated luminosity functions at $\bar{x}_{\rm HI}=0$ with different choices of the average mass of halos hosting LAEs ($\bar{M}_{\rm h}=2\times 10^{11}\text{, }2\times 10^{10}\text{ and }3\times 10^9M_\odot$ respectively, and $\beta=1$), which show a good agreement with the data.
For comparison we show the LAE luminosity function assuming $\bar{M}_{\rm h}=3\times 10^9M_\odot$ and $\beta=2/3$.
\label{fig:LAE}
}
\vspace{-1\baselineskip}
\end{figure}

\begin{figure}
\vspace{+0\baselineskip}
{
\includegraphics[width=0.45\textwidth]{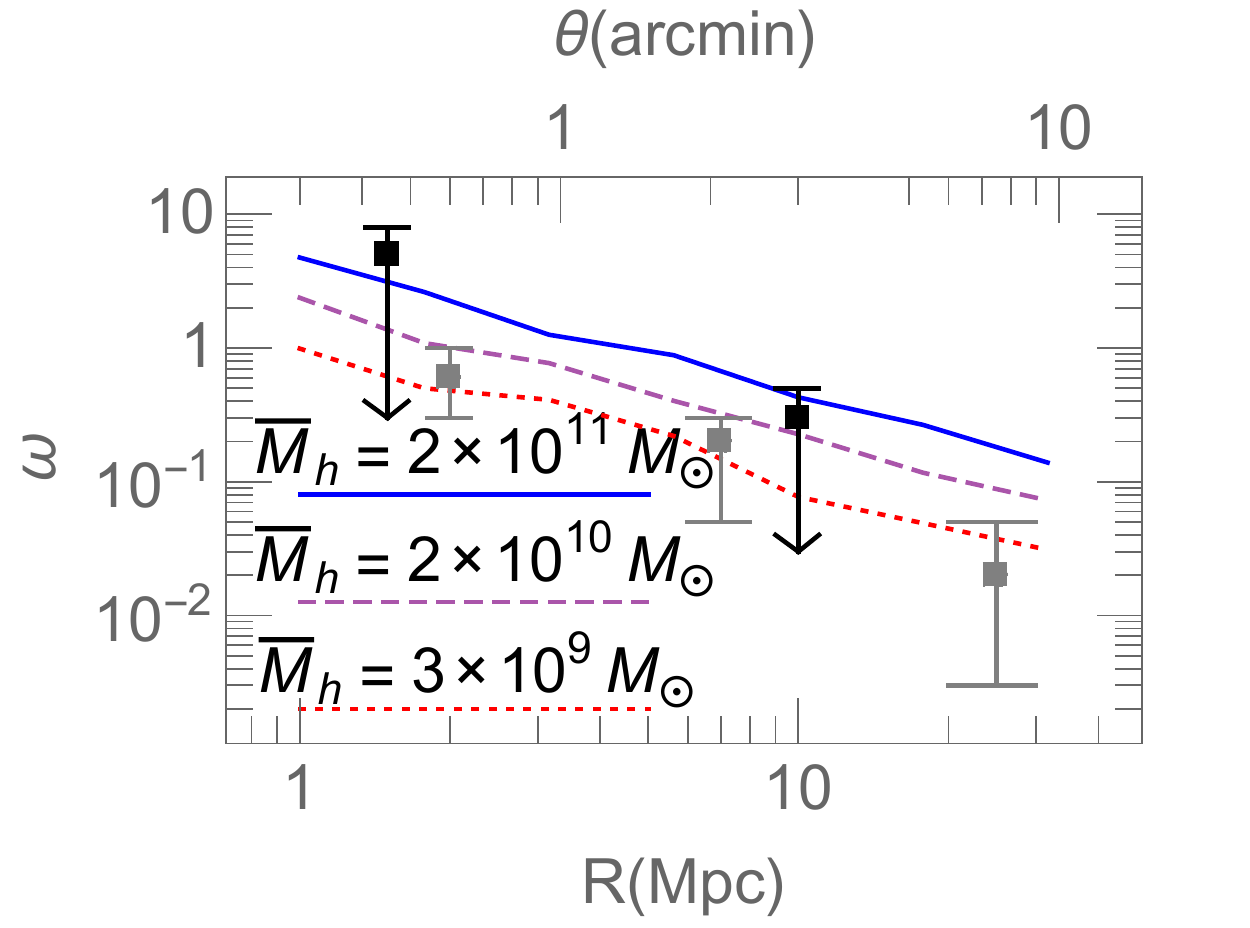}
}
\caption{Angular correlation functions of LAEs at z=6.6; black and grey dots represent observations by \citet{Ouchi10} for LAEs with $L_{\rm \alpha}^{\rm min}\simeq 4.0\times 10^{42}\text{ erg s$^{-1}$}$ and $L_\alpha^{\rm min}\simeq 2.5\times 10^{42}\text{ erg s$^{-1}$}$ respectively. With the solid, dashed, dotted lines we show the simulated correlation functions at $\bar{x}_{\rm HI}=0$ when $\bar{M}_{\rm h}=2\times 10^{11}\text{, }2\times 10^{10}\text{ and }3\times 10^9M_\odot$ respectively.
\label{fig:LAE_corr}
}
\vspace{-1\baselineskip}
\end{figure}

\subsubsection{Instrinsic line profiles}

The intrinsic profile of the \lya\ line is set by radiative transfer trough the ISM and CGM, and depends on dust, DLAs, geometry and gas kinematics (e.g. see the recent review by \citealt{Dijkstra14Review}).  All of these processes are highly uncertain in high-$z$ galaxies.  Here we merely evaluate the IGM absorption at a fixed velocity offset from the systemic redshift, chosen in our fiducial model to be $\Delta v =$200 km s$^{-1}$ (e.g. \citealt{Shibuya14, Stark15, Sobral15}).  We also show results using $\Delta v =$100 km s$^{-1}$ in Fig. A1, which serve to highlight that, unlike EoR constraints based on the (evolution of the) LAE number density,  {\it our large-scale clustering constraints are extremely insensitive to the chosen line profile}.  This is because we do not discriminate EoR models based on the observed LAE number density, $\nlea(z=6.6)$, as this quantity depends on both intrinsic and EoR properties.  As discussed above, we instead {\it evaluate clustering at a fixed number density} (see also \citealt{Jensen14}), effectively adjusting our free parameter $f_{\rm duty}$ to compensate for the stronger/weaker IGM absorption due to a smaller/larger systemic velocity offset.\footnote{This approach is different from the recent work of \citet{HDM15}, who instead fix $f_{\rm duty}=1$, and then use a line-centered Gaussian \lya\ profile in order to constrain how $\nlea$ changes during the EoR (which is in their model driven by the $z=6.6$ galaxies themselves).  With these added assumptions, they were able to obtain much tighter limits on $\avenf$ than we quote below.} 



\subsubsection{Comparison to current $z=6.6$ observations, assuming an ionized Universe}

We construct our mock survey by slicing the $z=6.6$ halo field of our simulations (see \S \ref{sec:sims}) into slabs of width 
$\simeq 40\text{ Mpc}$, corresponding to the redshift uncertainty, $\Delta z\simeq 0.1$, for the narrow-band LAE surveys \citep{Ouchi10}. Each halo is assigned an intrinsic \lya\ luminosity according to eq. (\ref{eq:LtoM}), and the IGM opacity is computed by stepping through the simulation from the halo center out to a distance of $\sim$ 200 Mpc along the chosen line of sight axis:
\begin{equation}
\tau_{\rm IGM}=\int dz \frac{cdt}{dz} n_{\rm H}\sigma_\alpha x_{\rm HI}\;,
\end{equation}
where $n_{\rm H}$ is the local hydrogen number density and $\sigma_\alpha$ is the Ly$\alpha$ cross-section  (taking into account proper motion of the cosmic gas).

Before discussing our reionization simulations, we first show LAE properties {\it assuming} $\avenf=0$ at $z=6.6$.\footnote{As discussed below, our EoR simulations have a sub-grid model for the percent-level residual fraction of HI inside ionized cells.  This means that $\tau_{\rm IGM}$ is not exactly zero post-reionization, and is slightly different in our two EoR models, owing to their different mass-weighted neutral fractions post-reionization ($\sim1$\% and $\sim3$\%).  These minor differences do not impact our conclusions, as any significant level of absorption post-reionization by proximate DLAs (e.g. \citealt{Mesinger15}), is by definition taken into account (on average) by our $L_{\alpha}^{\rm intr} \leftrightarrow M_{\rm h}$ relation.} All the quantities quoted below are calculated taking the average between all of the mock-survey slabs.

\begin{figure}
\vspace{+0\baselineskip}
{
\includegraphics[width=0.45\textwidth]{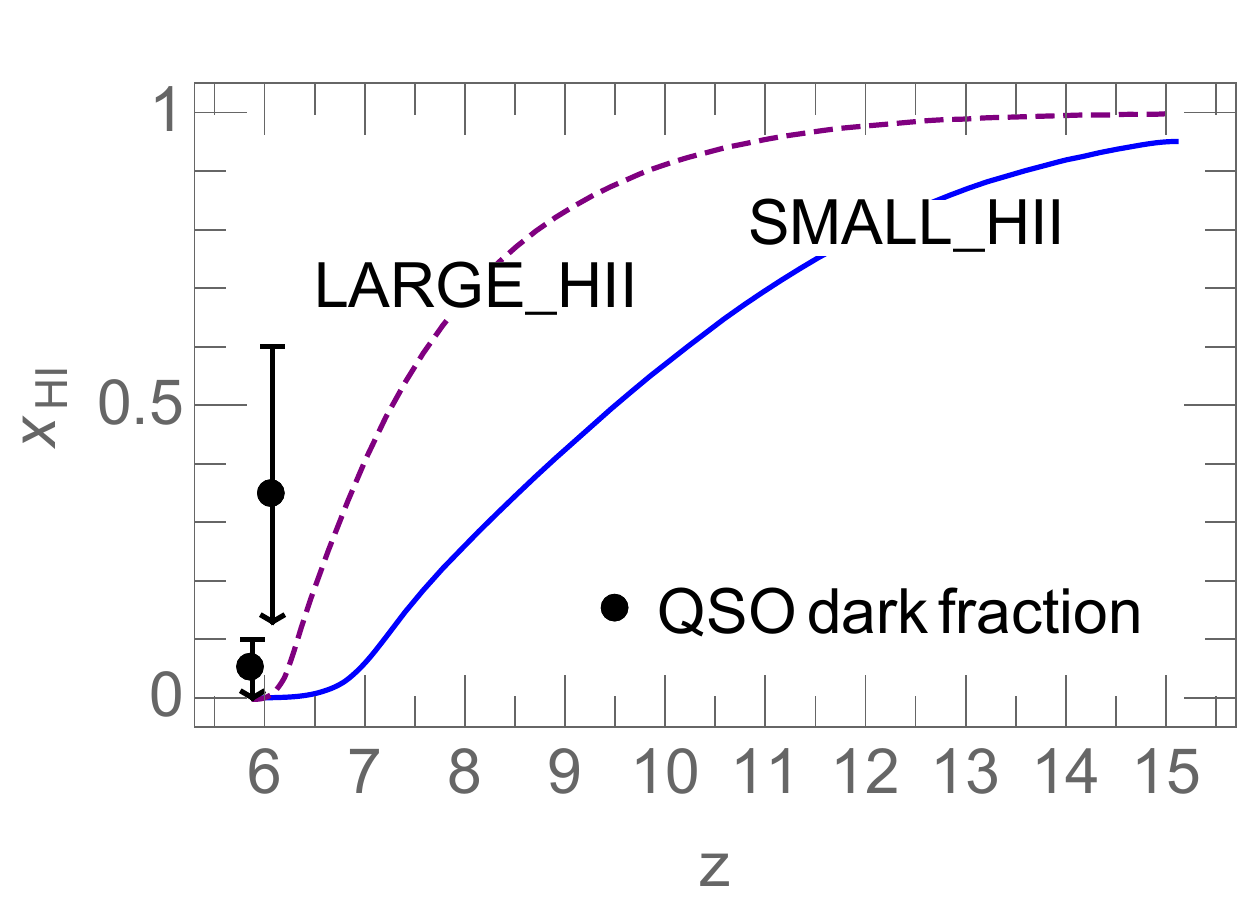}
}
\caption{Evolution of the volume-weighted global neutral fraction, $\bar{x}_{\rm HI}$ for the \textbf{SMALL\pmb{\_}HII} (solid) and \textbf{LARGE\pmb{\_}HII} (dashed) models; the \textbf{LARGE\pmb{\_}HII} model corresponds to EoR galaxies hosted only by the most biased halos, whereas the \textbf{SMALL\pmb{\_}HII} EoR model is driven by less biased, smaller-mass galaxies at the atomic cooling threshold.
This figure serves to illustrate physically-motivated values of $\avenf$ and its late-time evolution. In keeping with the systematic approach of this study however, we use the neutral fraction fields from various values of $\avenf$ when computing LAE clustering at $z\approx7$ (i.e. we do not assume a prior on $\avenf(z\approx7)$).
  We note that the \textbf{SMALL\pmb{\_}HII} (solid) and \textbf{LARGE\pmb{\_}HII} runs have $\tau_e=$ 0.07 and 0.05 respectively, both within the recent 1-$\sigma$ constraints published by Planck Collaboration 2015.  We also note on the plot  the strict, model-independent upper limits around $z\sim6$ from the dark fraction in QSO spectra (McGreer et al. 2015).
\label{fig:xHI_evo}
}
\vspace{-1\baselineskip}
\end{figure}

In Figure \ref{fig:LAE} we show the luminosity function of the LAEs at redshift $z=6.6$.  Dots represent observations by \citet{Ouchi10}, while diamonds represent the recent determination by \citet{Matthee15}. The curves correspond to our simulated luminosity functions.
With the lines we show our simulated luminosity functions assuming $\bar{x}_{\rm HI}=0$.  The assumption that $L_{\alpha}^{\rm intr}\propto M_{\rm h}$(i.e. $\beta=1$) is at odds with the recent observations by \citet{Matthee15} at the bright end, {\it if} LAEs are hosted by the most massive halos ($f_{\rm duty}\approx1$).  Although one could flatten the LFs by assuming a weaker dependence on halo mass (lower $\beta$), we show below that this extreme {\it most massive halos} model is ruled out with clustering measurements.  Although here we assume a post-reionization Universe, we note that reionization should not appreciably change the {\it shape} of the observed LFs, only its amplitude (e.g. \citealt{FZH06, McQuinn07LAE, MF08LAE, DMF11}).

In Figure \ref{fig:LAE_corr} we show the LAE ACFs at $z=6.6$, again assuming $\bar{x}_{\rm HI}=0$. Black and grey dots represent observations by \citet{Ouchi10} for LAEs with $L_\alpha^{\rm min}\simeq 4.0\times 10^{42}\text{ erg s$^{-1}$}$ and $L_\alpha^{\rm min}\simeq 2.5\times 10^{42}\text{ erg s$^{-1}$}$ respectively. With the solid, dashed and dotted lines we show the simulated ACF for our three intrinsic luminosity relations: $\bar{M}_{\rm h}=2\times 10^{11}\text{, }2\times 10^{10}\text{ and }3\times 10^9M_\odot$ respectively. We point out that, since LAEs with $L_{\alpha}^{\rm intr}\gtrsim 4.0\times 10^{42}\text{ erg s$^{-1}$}$ provide only upper limits on the ACF, they can not be used to discriminate models with different $\bar{M}_{\rm h}$; instead LAEs with a lower luminosity $L_{\alpha}^{\rm intr}\gtrsim 2.5\times 10^{42}\text{ erg s$^{-1}$}$ provide useful constraints on $\bar{M}_{\rm h}$, and exclude the most extreme $\bar{M}_{\rm h}=2\times 10^{11}M_\odot$, as already noted by \citet{Ouchi10}.
Here we are assuming $\bar{x}_{\rm HI}=0$; however, a higher neutral fraction increases the ACF at the same $\bar{M}_{\rm h}$, strengthening this conclusion.

\subsection{Large-scale simulations of reionization}
\label{sec:sims}

\begin{figure*}
\vspace{+0\baselineskip}
{
  \includegraphics[width=0.33\textwidth]{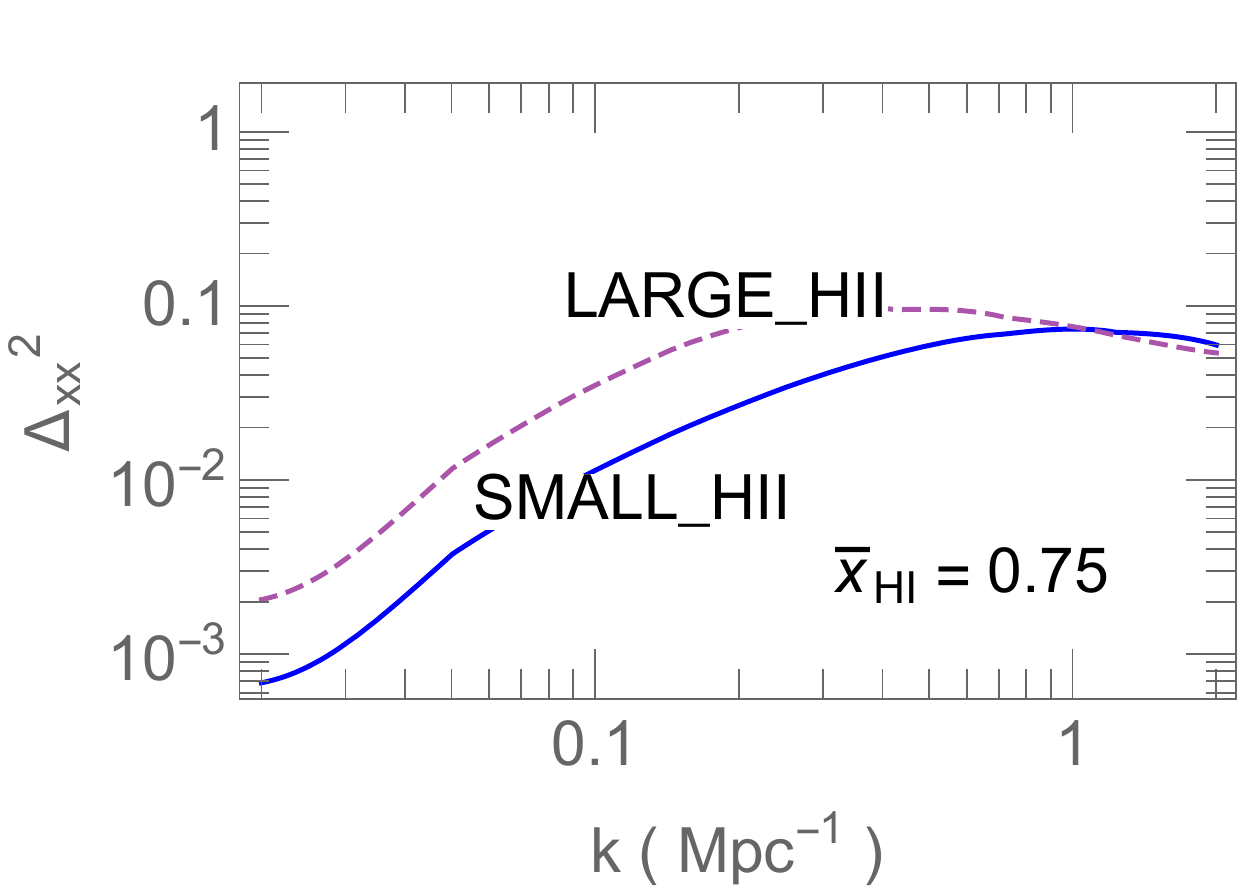}
  \includegraphics[width=0.33\textwidth]{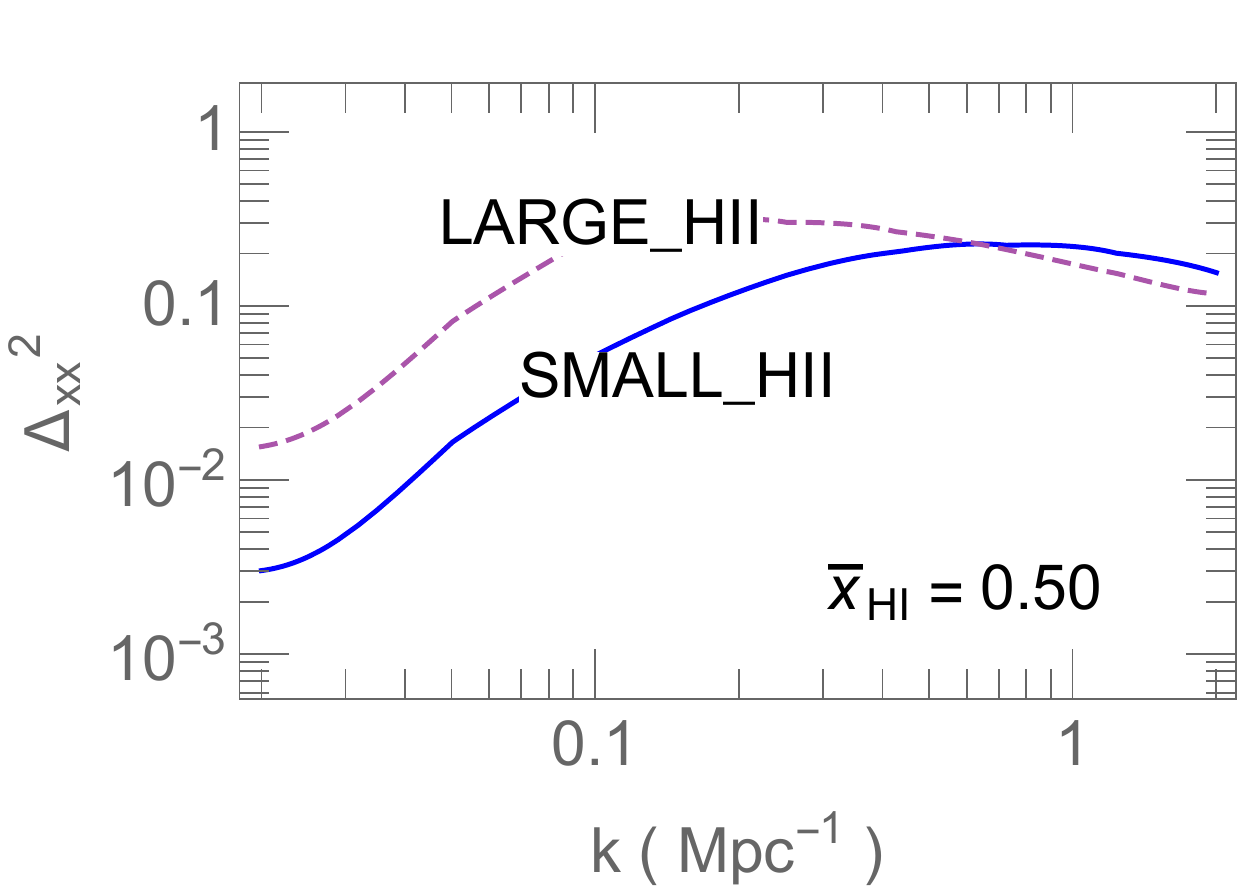}
  \includegraphics[width=0.33\textwidth]{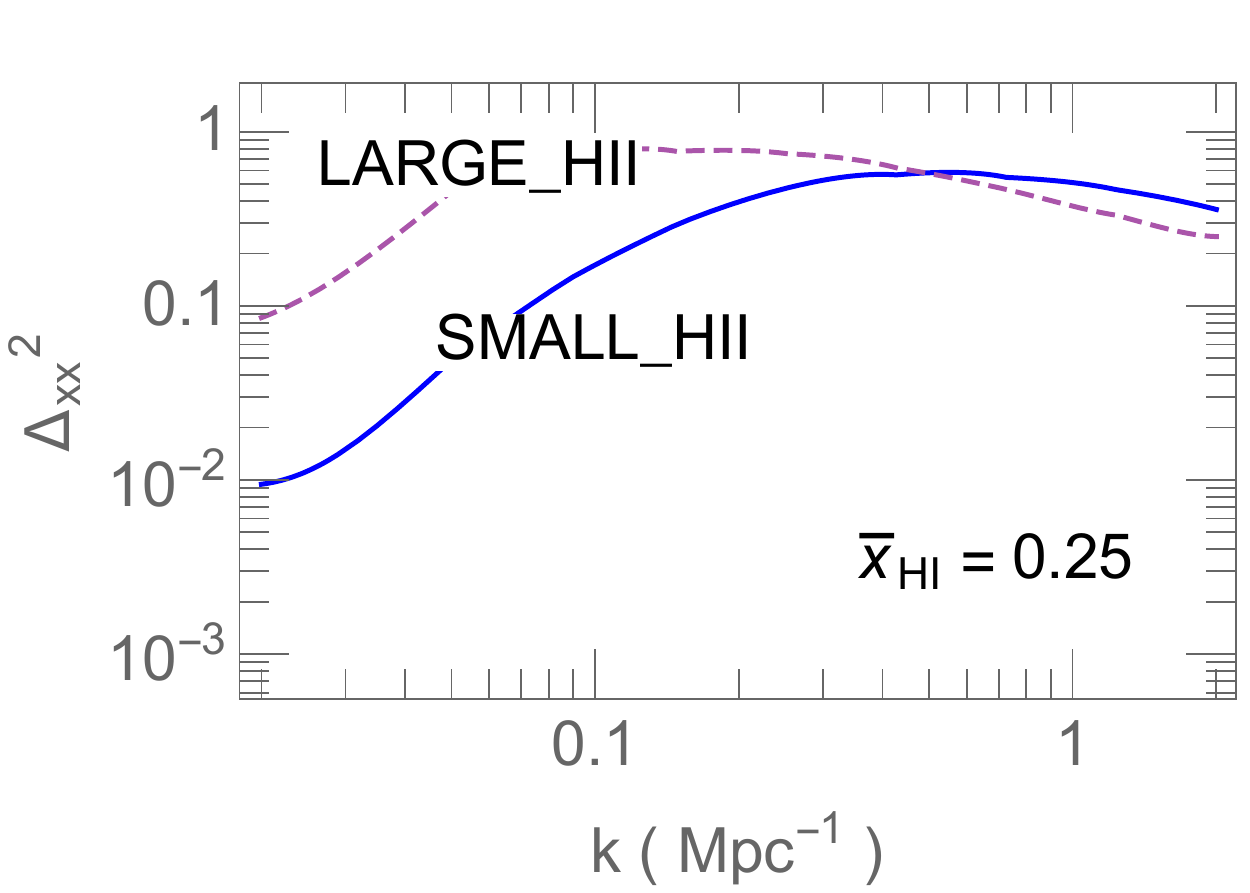}
}
\caption{Power spectra of the ionization fields at $\bar{x}_{\rm HI}=$ 0.75, 0.50, 0.25 (left to right panels) for the \textbf{SMALL\pmb{\_}HII} (solid) and \textbf{LARGE\pmb{\_}HII} (dashed) EoR models.
\label{fig:pow_spec}
}
\vspace{-1\baselineskip}
\end{figure*}

To model cosmological reionization, we use \cmfastSM\ \citep{SM13b, SM14}, a modified version of the publicly-available code \cmfast\footnote{http://homepage.sns.it/mesinger/Sim} \citep{MF07, MFC11}.  We briefly describe the main features of the code below; interested readers are encouraged to read the introductory papers.

We generate the IGM density and source fields by: (i) creating a 3D Monte Carlo realization of the linear density field in a box with sides $L=500\text{ Mpc}$ and $N=2500^{3}$ grid cells; (ii) evolving the density field using the Zeldovich approximation \citep{Zeldovich70}, and smoothing onto a lower-resolution $N=500^3$ grid; (iii) using excursion-set theory \citep{PS74, LC93, ST99} on the evolved density field to compute the fraction of matter collapsed in halos bigger than a threshold halo mass, assumed to be large enough to host star-forming galaxies (see below).  The Lagrangian positions of $z\approx7$ halos are identified according to the standard spherical collapse model, and subsequently displaced using the Zeldovich approximation\footnote{We confirm that using second order Lagrangian perturbation theory only changes the resulting $z\approx7$ correlation functions by $\leq$ 2-5\%, on the scales considered in this work.} \citep{MF07}.

The ionization field is computed by comparing the cumulative number of ionizing photons to the number of baryons plus cumulative recombinations, in spherical regions of decreasing radius $R$ (i.e. following the excursion-set approach of \citealt{FHZ04}).
Specifically, a cell located at spatial position and redshift, ($\textbf{x}$, $z$), is flagged as ionized if:
\begin{equation}
\label{eq:ion_crit_coll}
\xi f_{\rm coll}(\textbf{x}, z, R, \bar{M}_{\rm min})\geq 1+\bar{n}_{\rm rec}(\textbf{x}, z, R)
\end{equation}
where the ionizing efficiency can be written as:
\begin{equation}
\label{eq:zeta}
\xi = 30 \bigg(\frac{N_\gamma}{4000}\bigg) \bigg(\frac{f_{\rm esc}}{0.15}\bigg) \bigg(\frac{f_\ast}{0.05}\bigg) \bigg(\frac{f_{\rm b}}{1}\bigg)~ ,
\end{equation}
and $N_\gamma$ is the number of ionizing photons per stellar baryon, $f_{\rm esc}$ is the fraction of UV ionizing photons that escape into the IGM, $f_\ast$ is the fraction of galactic gas in stars, $f_{\rm b}$ is the fraction of baryons inside the galaxy with respect to the cosmic mean $\Omega_{\rm b}/\Omega_{\rm m}$, and $f_{\rm coll}\left(\textbf{x},z, R, \bar{M}_{\rm min}\right)$ is the fraction of collapsed matter inside a sphere of radius $R$ residing in halos larger than $\bar{M}_{\rm min}$. \footnote{Note that the minimum halo mass of cosmic reionization sources, $M_{\rm min}$, is a different quantity than the halo mass corresponding to the faintest $z=6.6$ LAEs discussed above, $M_{\rm \alpha}^{\rm min}$.} Starting from the box size, the smoothing scale is decreased (averaging also over $M_{\rm min}($\textbf{x}$, $z$)$ and $n_{\rm rec}($\textbf{x}$, $z$)$), and the criterion in eq. (\ref{eq:ion_crit_coll}) is re-evaluated to determine if a cell is ionized or not.  For the remaining neutral cells, the partial ionizations from sub-grid sources are accounted for by setting the ionized fraction of the remaining neutral cells to $\xi f_{\rm coll}({\bf x}, z, R_{\rm cell}, M_{\rm min})/({1+n_{\rm rec}})$ \citep{Zahn11}.

As detailed in \citet{SM14}, the main improvement of \cmfastSM\ is for each cell to keep track of its local values of $M_{\rm min}($\textbf{x}$, $z$)$ and $n_{\rm rec}($\textbf{x}$, $z$)$, computed according to the cell's density and ionization history. These values depend on density and ionization structure on scales much smaller than the cell sizes of EoR simulations, and therefore must be accounted for in a sub-grid fashion.

The minimum halo mass hosting star-forming galaxies can be expressed as:
\begin{equation}
\label{eq:M_min}
M_{\rm min}=\max\left[M_{\rm cool}, M_{\rm crit}\right] ~,
\end{equation}
where $M_{\rm cool}(z)$ corresponds to the atomic cooling threshold of halos with virial temperatures $\gsim10^4$ K, and $M_{\rm crit}(J_{21}, z_{\rm IN}, z)$ corresponds to a feedback scale below which star-formation is very inefficient in regions exposed to a UV background with intensity $J_{21}$ since $z_{\rm IN}$ (see below). 
Likewise, the local cumulative number of recombinations per baryon is computed according to:
\begin{equation}
\label{eq:n_rec}
n_{\rm rec}(\textbf{x}, z)=\int_{z_{\rm IN}}^{z}\frac{dn_{\rm rec}}{dt}\frac{dt}{dz}dz ~ ,
\end{equation}
with
\begin{equation}
\label{eq:dn_rec}
\frac{d n_{\rm rec}}{dt}(\textbf{x}, z)=\int_0^{+\infty}P_{\rm V}\left(\Delta, z\right)\Delta\bar{n}_{\rm H}\alpha_{\rm B}\left[1-x_{\rm HI}\left(\Delta\right)\right]^{2}d\Delta ~,
\end{equation}
\noindent where $P_{\rm V}\left(\Delta, z\right)$ is the full (non-linear, sub-grid) distribution of overdensities: $\Delta\equiv n_{\rm H}/\bar{n}_{\rm H}$ (e.g. \citealt{MHR00, SM14}), and  $x_{\rm HI}\left(\Delta\right)$ is equilibrium neutral fraction computed using the empirical self-shielding prescription of \citet{Rahmati13}.

\subsubsection{EoR Runs}

Given the current uncertainties on the reionization history and morphology, we compare the results of two different runs, which should span the range of possible EoR morphologies:
\begin{itemize}
\item \textbf{SMALL\pmb{\_}HII}: corresponding to the fiducial simulation in \citet{SM14}, this run ignores internal feedback\footnote{Supernovae can blow out gas from small potential wells, thereby dramatically reducing subsequent star formation inside the smallest halos (e.g.  \citealt{SH03, PS09, FDO11}).  The efficiency of such internal feedback and what precisely is the associated suppression scale, depend on the details of SNe explosions, cooling, and ISM structure, and are extremely uncertain at high-redshifts.}
with only photo-heating feedback suppressing star formation. Specifically, we take $\xi=30$ and $M_{\rm crit}=M_{\rm 0}J_{\rm 21}^{\phantom{21}a}\left(\frac{1+z}{10}\right)^{b}\left[1-\left(\frac{1+z}{1+z_{\rm IN}}\right)^{c}\right]^{d}$, with best-fit values from \citet{SM13a}: $\left(M_{\rm 0}, a, b, c, d\right)=\left(2.8\times 10^{9}M_{\odot}, 0.17, -2.1, 2.0, 2.5\right)$.  The negative impact of photo-heating feedback on the baryon budget at $z\gsim6$ is modest on its own (e.g. \citealt{OGT08, SM13b, NM14, PSV15}).  However, its importance is dramatically increased when inhomogeneous recombinations are taken into account \citep{SM14}, as both processes preferentially slow the expansion of the largest HII regions, which started growing early-on, thus allowing sufficient time for both feedback processes to take their toll.  The resulting EoR models are extended and {\it characterized by small cosmic HII patches}.

\item \textbf{LARGE\pmb{\_}HII}: this run assumes that star-formation is efficient only in the rarest, most-massive, most-biased halos.  Such a model could be motivated by efficient internal feedback inside small mass galaxies (e.g. \citealt{SH03, RTL11, Kim13}). We adopt the extreme values of $\xi=50$ and $M_{\rm crit}(T_{\rm vir}=10^5\text{ K})$.  Such a high threshold virial temperature approximately corresponds to (a factor of $\sim$2 lower than) that inferred from the faintest observed high-$z$ galaxies, as computed from abundance matching (e.g. \citealt{KF-G12, GM15}). Driven by rapidly-forming galaxies on the high-mass tail of the mass function, the resulting EoR models are rapid and {\it characterized by large cosmic HII patches}.
\end{itemize}

In Figure \ref{fig:xHI_evo} we compare the evolution of the global volume-averaged neutral fraction $\bar{x}_{\rm HI}$ in these two different reionization models. This serves only to illustrate the plausible range of EoR evolutions: having a mostly neutral Universe at $z=6.6$ is difficult to achieve with physical EoR models (see also, e.g. \citealt{Lidz07}).  However,  in keeping with the systematic nature of our study, {\it we do not restrict the allowed values of $\avenf$ at $z\approx7$}, instead computing the resulting LAE clustering with ionization maps at any $\avenf(z)$ superimposing them on the $z\approx7$ halo field (e.g. \citealt{McQuinn07LAE, MF08LAE}).

Therefore, much more relevant to this work is the EoR morphology of the two models, which is quantified in Fig. \ref{fig:pow_spec}.  We plot the spherically-averaged power spectra, $\Delta^2_{\rm xx} \equiv k^3/(2\pi^2 V) ~ \langle|\delta_{\rm xx}|^2\rangle_k$, with $\delta_{\rm xx}=x_{\rm HI}/\avenf - 1$ at $\bar{x}_{\rm HI}=$ 0.75, 0.5, 0.25.  As expected, the resulting EoR morphology is very different in the two models, with \textbf{LARGE\pmb{\_}HII} having $\sim$2--5 times more power than \textbf{SMALL\pmb{\_}HII} on large ($k\lsim0.1$ Mpc$^{-1}$) scales.

Finally, we note that, unlike recent work on the \lya\ fraction in \citet{Mesinger15}, we do not resolve the opacity of self-shielded systems.
 Although originally thought to be important in governing the \lya\ transmission at high-redshifts \citep{BH13}, their impact is dramatically diminished with more realistic self-shielding prescriptions (\citealt{Mesinger15}, see also \citealt{Keating14}).  Moreover, due to their  steep damping wing profiles (e.g. \citealt{Miralda-Escude98, MF08damp}), self-shielded systems have to be proximate to the galaxy in order to have a strong imprint on the \lya\ line.  This means that their potential impact on the detectability of LAEs can be mostly subsumed within the uncertainties in modeling the $L_{\alpha}^{\rm intr} \leftrightarrow M_{\rm h}$ relation discussed above, and is highly unlikely to have a strong imprint on the clustering signature on the $\sim$10 Mpc scales relevant for the EoR (e.g. \citealt{Crociani11}).

\section{Results}
\label{sec:results}

\subsection{Mock LAE maps}

\begin{figure*}
\vspace{+0\baselineskip}
{
\includegraphics[width=0.28\textwidth, height=0.32\textwidth]{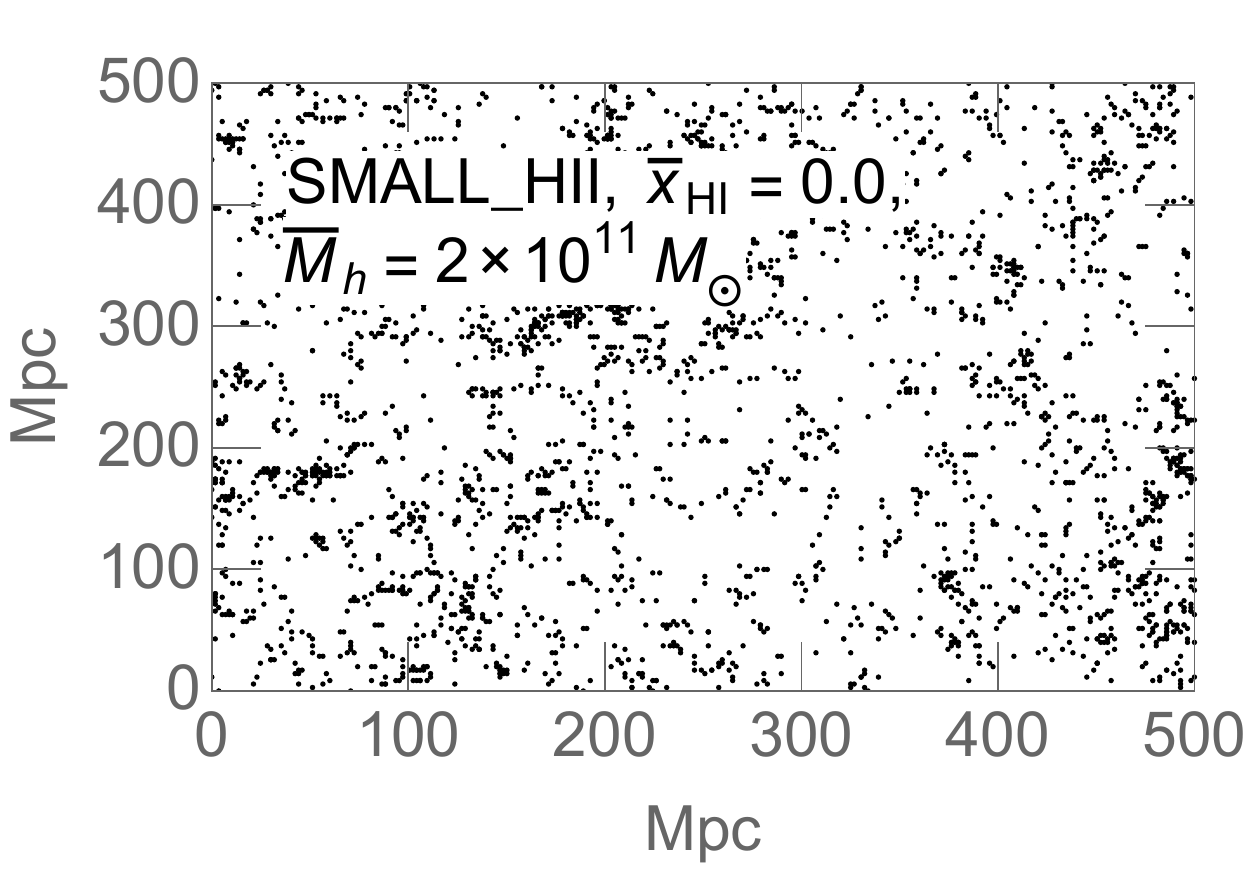}\qquad
\includegraphics[width=0.28\textwidth, height=0.32\textwidth]{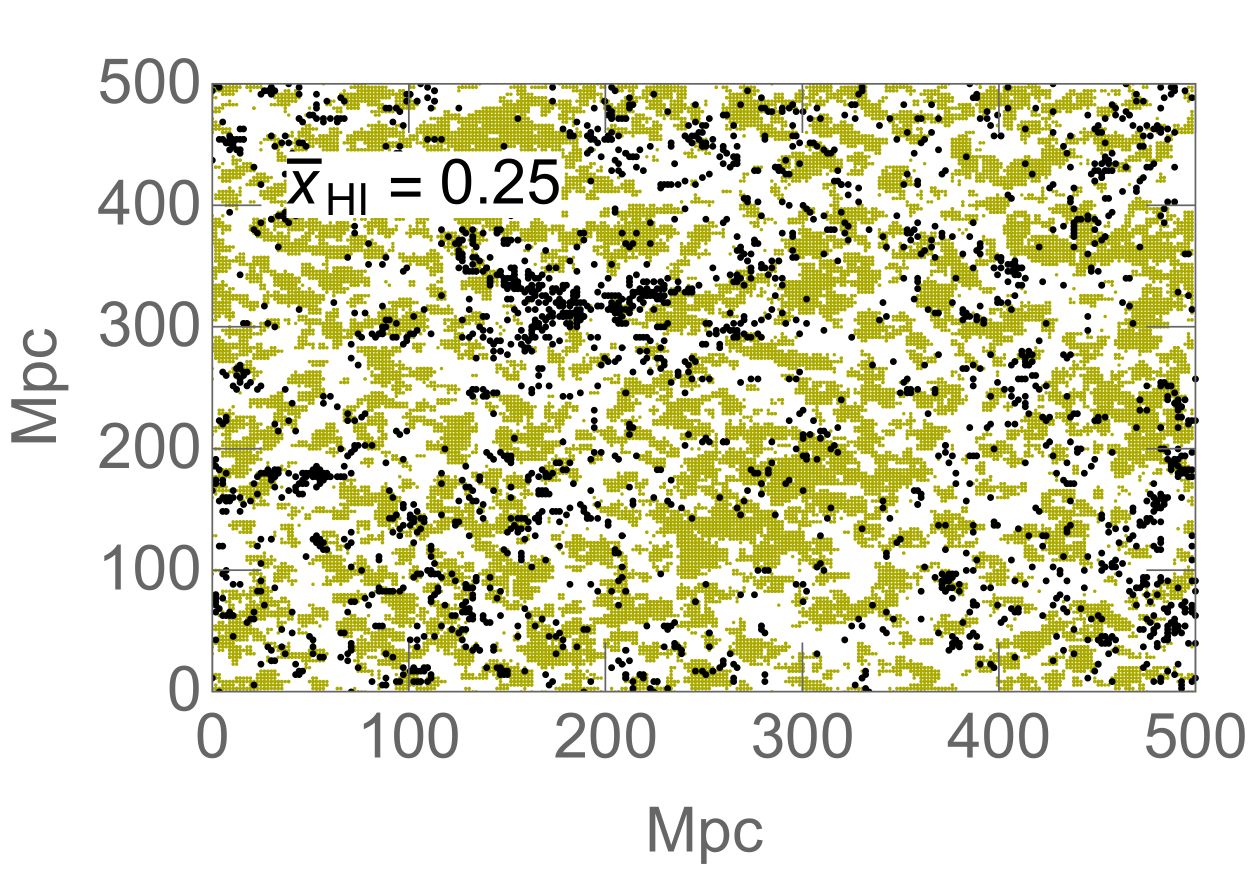}\qquad
\includegraphics[width=0.28\textwidth, height=0.32\textwidth]{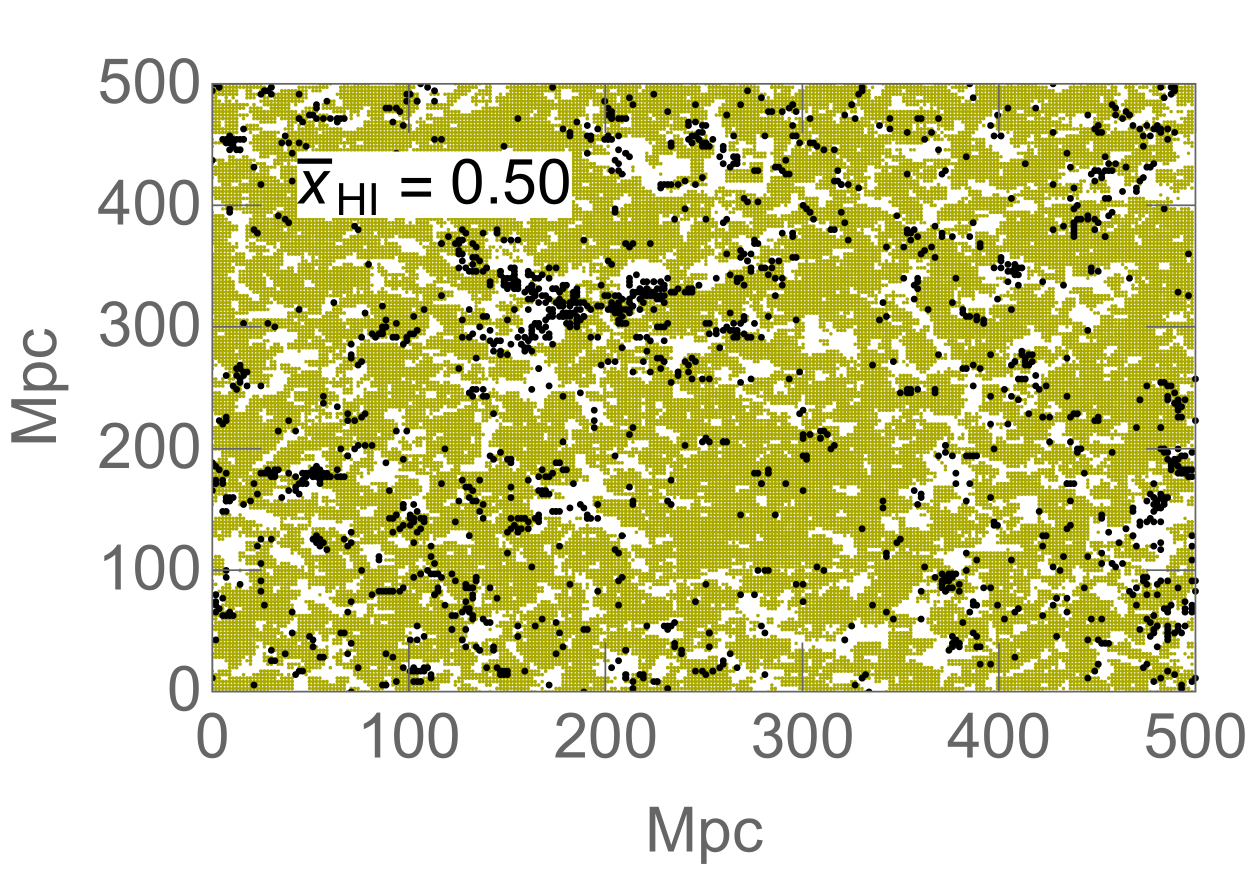}
\includegraphics[width=0.28\textwidth, height=0.32\textwidth]{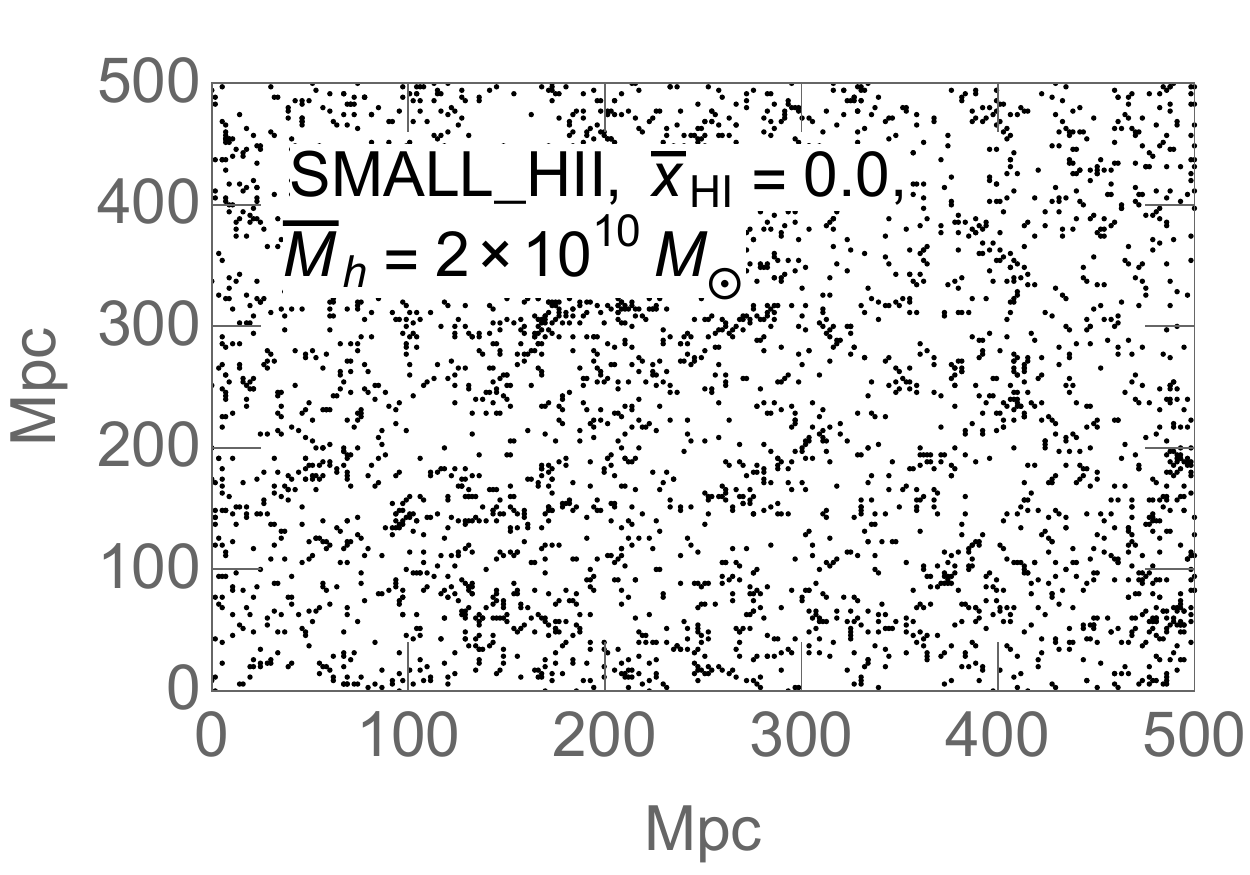}\qquad
\includegraphics[width=0.28\textwidth, height=0.32\textwidth]{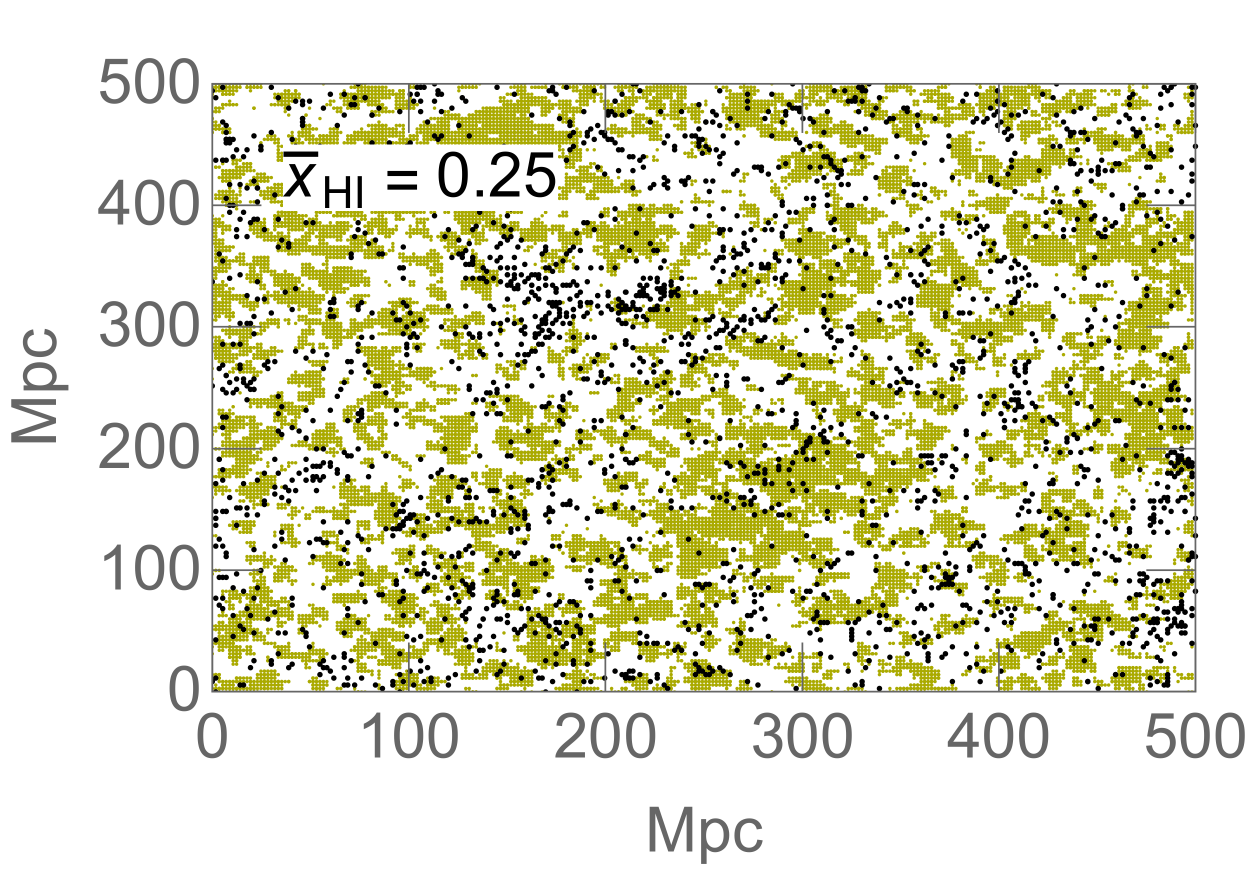}\qquad
\includegraphics[width=0.28\textwidth, height=0.32\textwidth]{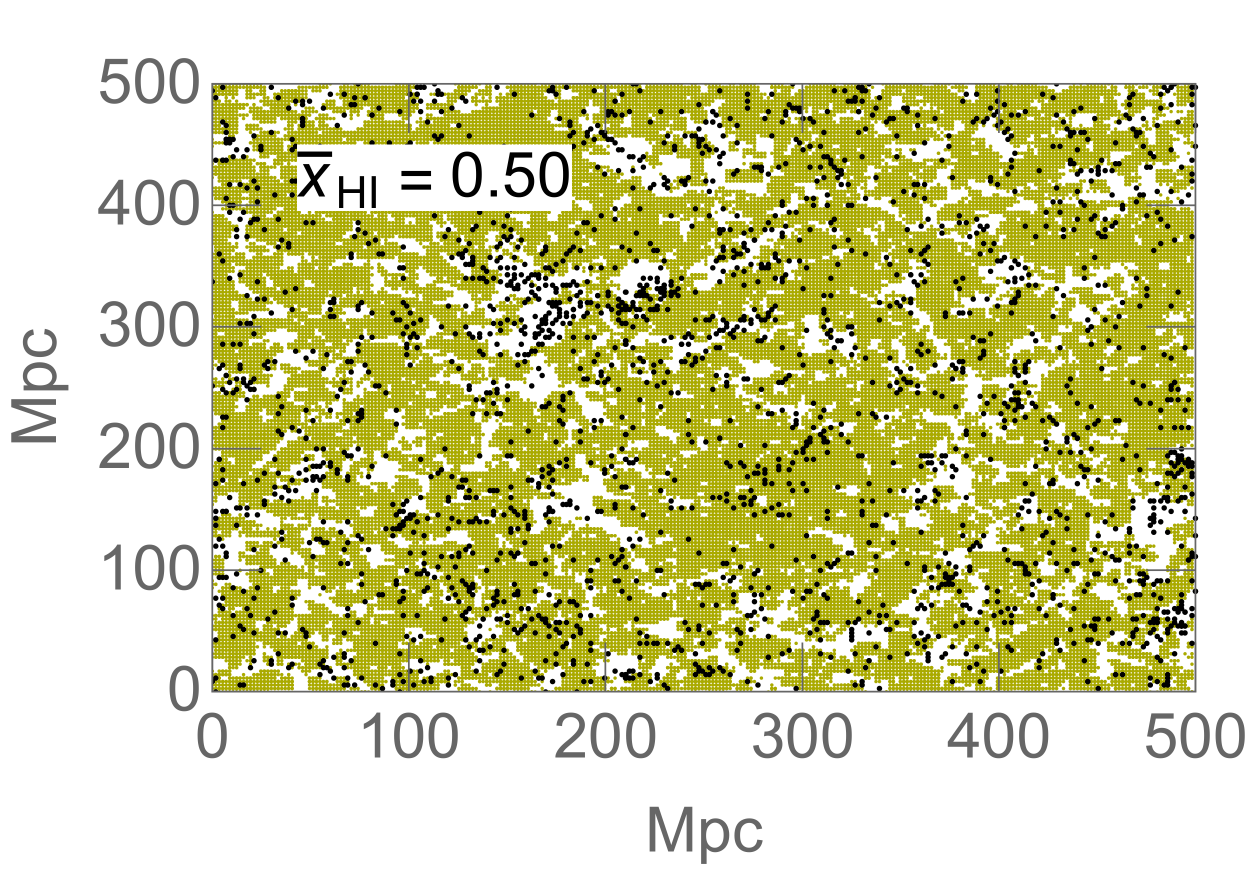}
\includegraphics[width=0.28\textwidth, height=0.32\textwidth]{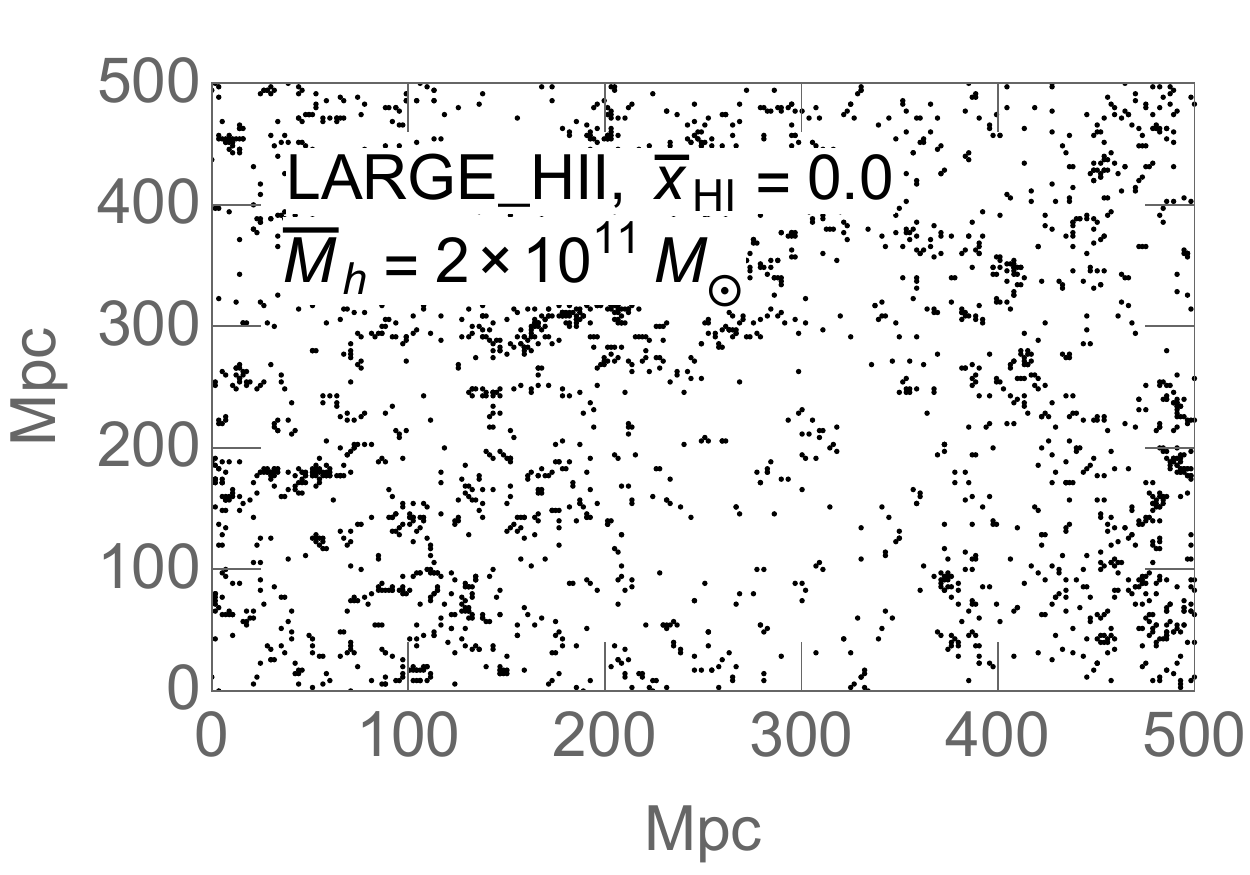}\qquad
\includegraphics[width=0.28\textwidth, height=0.32\textwidth]{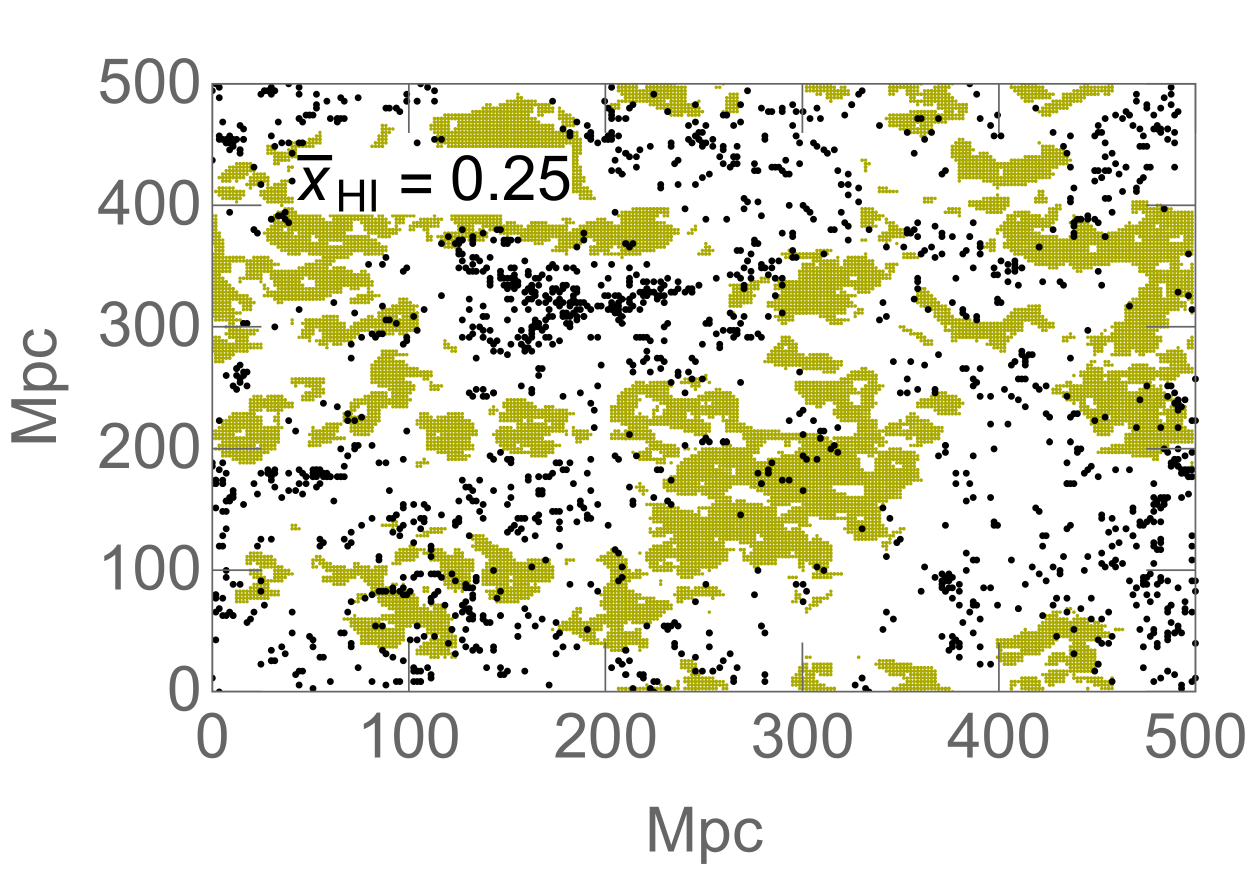}\qquad 
\includegraphics[width=0.28\textwidth, height=0.32\textwidth]{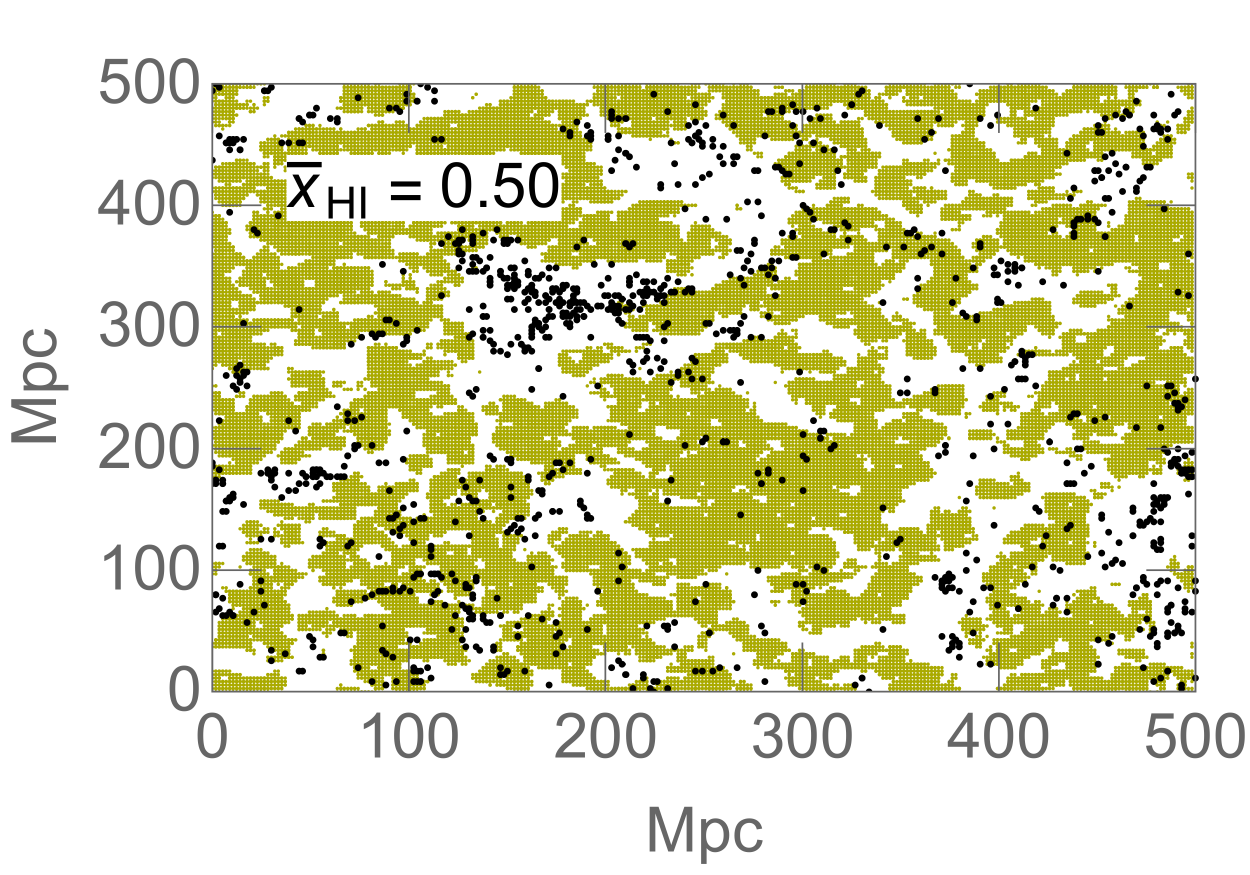}
\includegraphics[width=0.28\textwidth, height=0.32\textwidth]{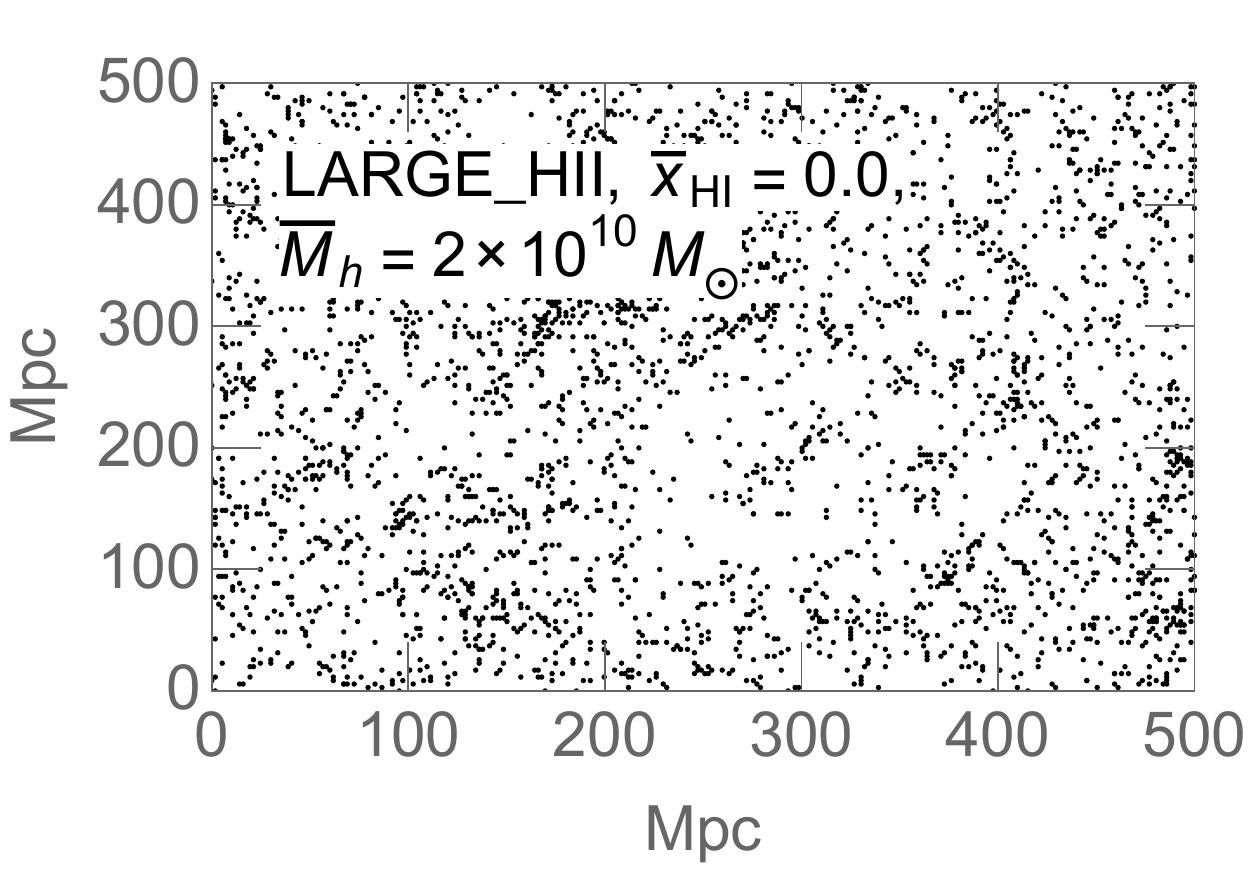}\qquad
\includegraphics[width=0.28\textwidth, height=0.32\textwidth]{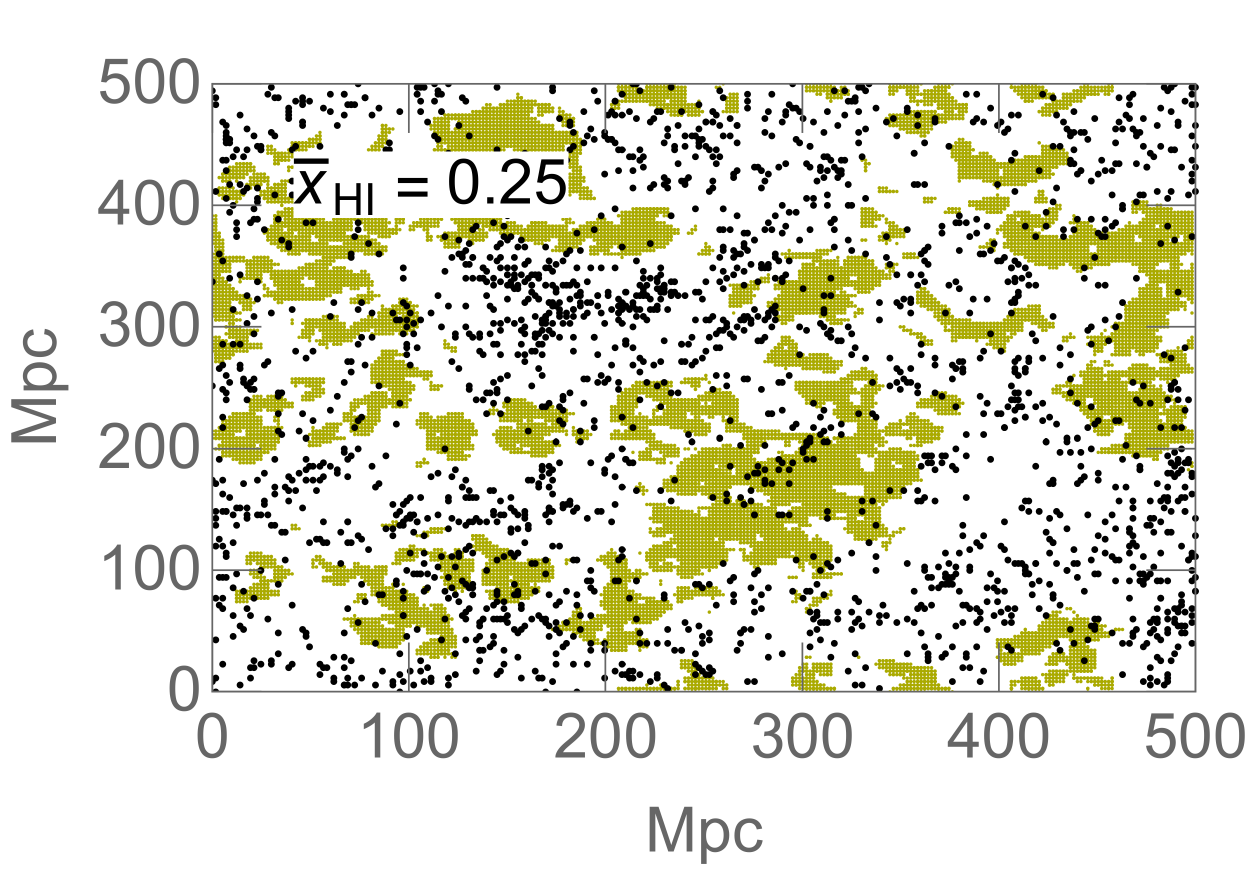}\qquad
\includegraphics[width=0.28\textwidth, height=0.32\textwidth]{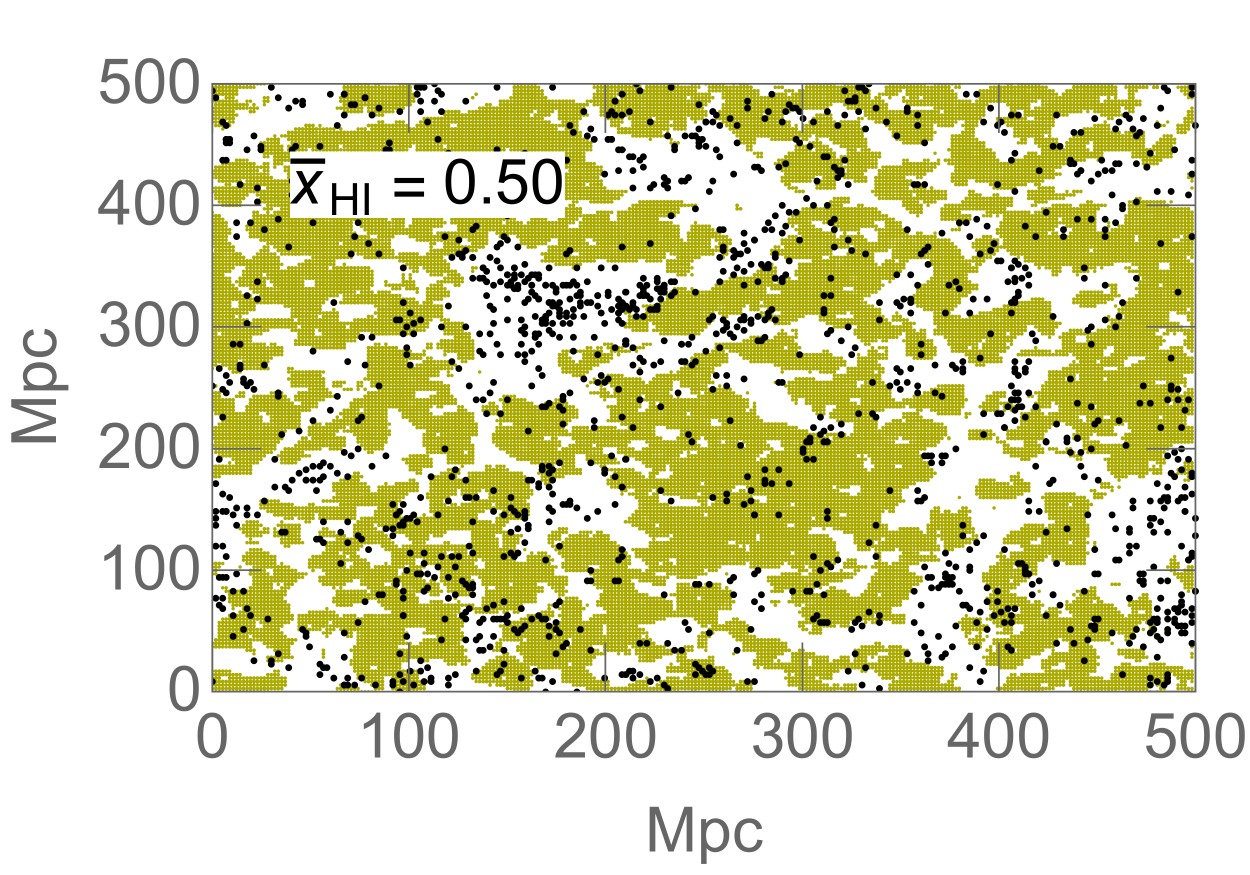}
}
\caption{Distribution of the observed LAEs in our mock surveys at different ionization fractions ($\bar{x}_{\rm HI}=0$, $\bar{x}_{\rm HI}=0.25$, $\bar{x}_{\rm HI}=0.50$ from left to right); the panels show different reionization models and $L_{\alpha}^{\rm intr} \leftrightarrow M_{\rm h}$ prescriptions: \textbf{SMALL\pmb{\_}HII}, $\bar{M}=2\times 10^{11}M_\odot$; \textbf{SMALL\pmb{\_}HII}, $\bar{M}=2\times 10^{10}M_\odot$; \textbf{LARGE\pmb{\_}HII}, $\bar{M}=2\times 10^{11}M_\odot$; \textbf{LARGE\pmb{\_}HII}, $\bar{M}=2\times 10^{10}M_\odot$ ({\it top to bottom}). All of the surveys have the same observed number density of LAEs $\bar{n}_{\rm LAE}=4.1\times10^{-4}$ Mpc$^{-3}$.
\label{fig:LAE_distr}
}
\vspace{-1\baselineskip}
\end{figure*}

In Figure \ref{fig:LAE_distr} we show the distribution of the observed LAEs (black points) in our mock surveys.  Columns correspond to increasing neutral fractions: $\bar{x}_{\rm HI}=$ 0, 0.25, 0.5 (left to right), with 1 Mpc thick slices through the reionization fields overlaid over the observed LAEs (green corresponds to neutral patches).  Rows correspond to the different $L_{\alpha}^{\rm intr} \leftrightarrow M_{\rm h}$ prescriptions and different reionization morphologies, as indicated in the labels.  As described above, all mock surveys are shown at a fixed observed number density of LAE, $\bar{n}_{\rm LAE}=4.1\times10^{-4}$ Mpc$^{-4}$ \citep{Ouchi10}, integrated along the survey depth of $\Delta z \approx 0.1$. This requires a higher LAE duty cycle earlier in reionization (for a given $\bar{M}_{\rm h}$), in order to compensate for the additional attenuation of the LAEs during the EoR without changing their intrinsic clustering.

There are several trends which can already be seen in Fig. \ref{fig:LAE_distr}.  Firstly, there is an increase in the clustering of observed LAE in the early stages of reionization \citep{McQuinn07LAE, MF08LAE, Jensen13, HDM15}, which is evident even when the maps are compared at a fixed observed number density.
However, the clustering seems insensitive to the reionization morphology, at fixed $\avenf$ (e.g. \citealt{McQuinn07LAE}).  This is mostly due to the fact that the narrow band surveys only localize the LAE within $\Delta z\approx0.1$  (corresponding to $\sim 40\text{ Mpc}$ at $z=6.6$), which is much larger than the typical scale of cosmic HII regions of EoR models.  Therefore the information on the reionization morphology is lost when smoothing over the large redshift depth of these narrow band surveys.  This can also be seen in the figure by noting that the positions of the observed LAEs only weakly correlate with the thin, 1Mpc slices through the neutral fraction fields shown in green.

Secondly, we see from Fig. \ref{fig:LAE_distr} that the typical mass of halos hosting LAEs also has a notable imprint on their clustering.  Given that we do not understand LAEs at high-$z$, this poses both a challenge (in disentangling the clustering signature from the one imposed by patchy reionization), and an opportunity (in allowing us to better understand \lya\ emission from galaxies using the large number statistics from upcoming surveys).

To sum up, the observed clustering at a fixed LAE number density should be strongly impacted by: (i) the typical halo masses hosting LAEs, $\bar{M}_{\rm h}$; and (ii) the neutral fraction of the IGM, $\avenf$.  The clustering is far less affected by the reionization model (i.e. morphology of reionization).  In the next section we quantify these statements.

\subsection{Angular correlation functions}

\begin{figure*}
\vspace{+0\baselineskip}
{
\includegraphics[width=0.45\textwidth]{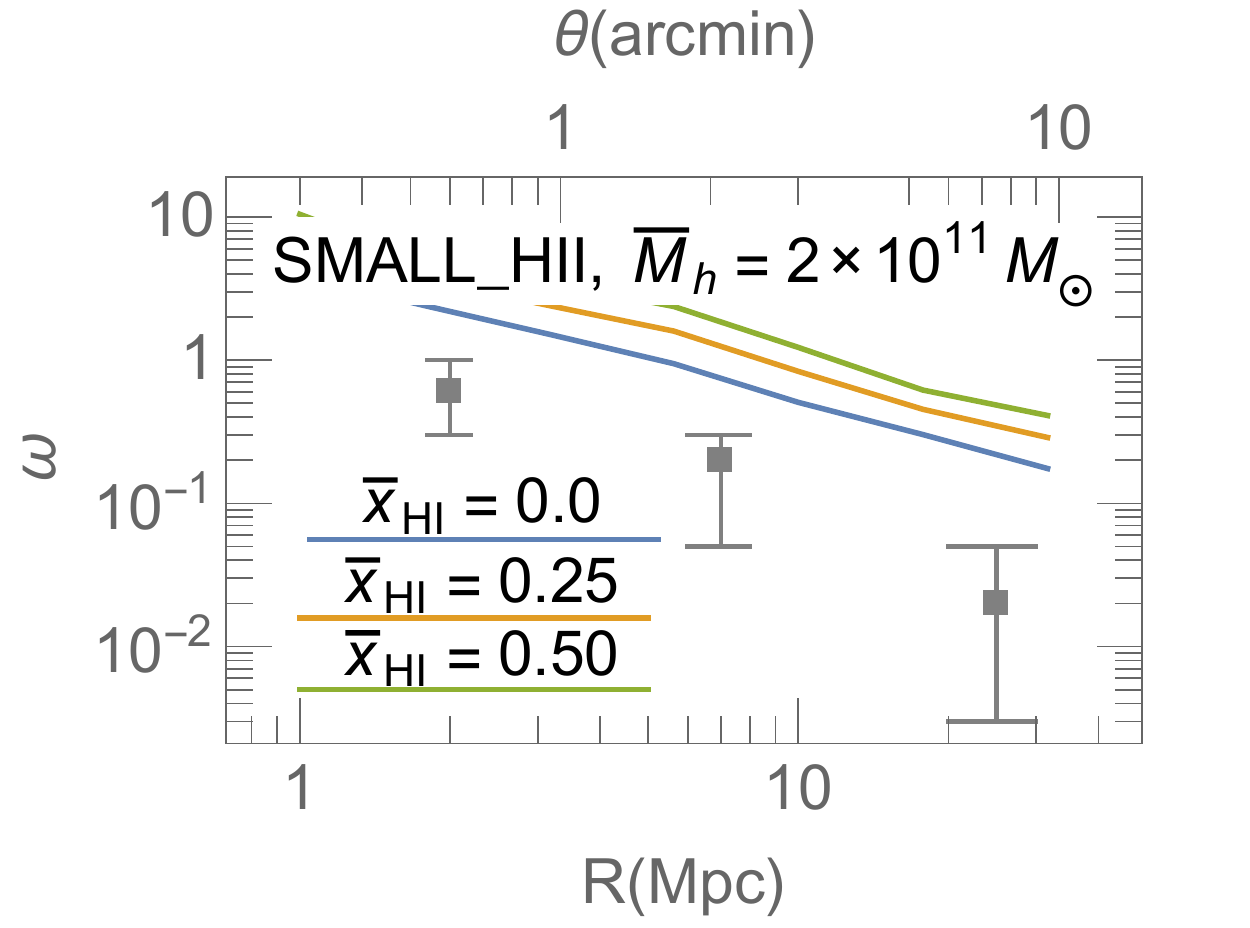} 
\includegraphics[width=0.45\textwidth]{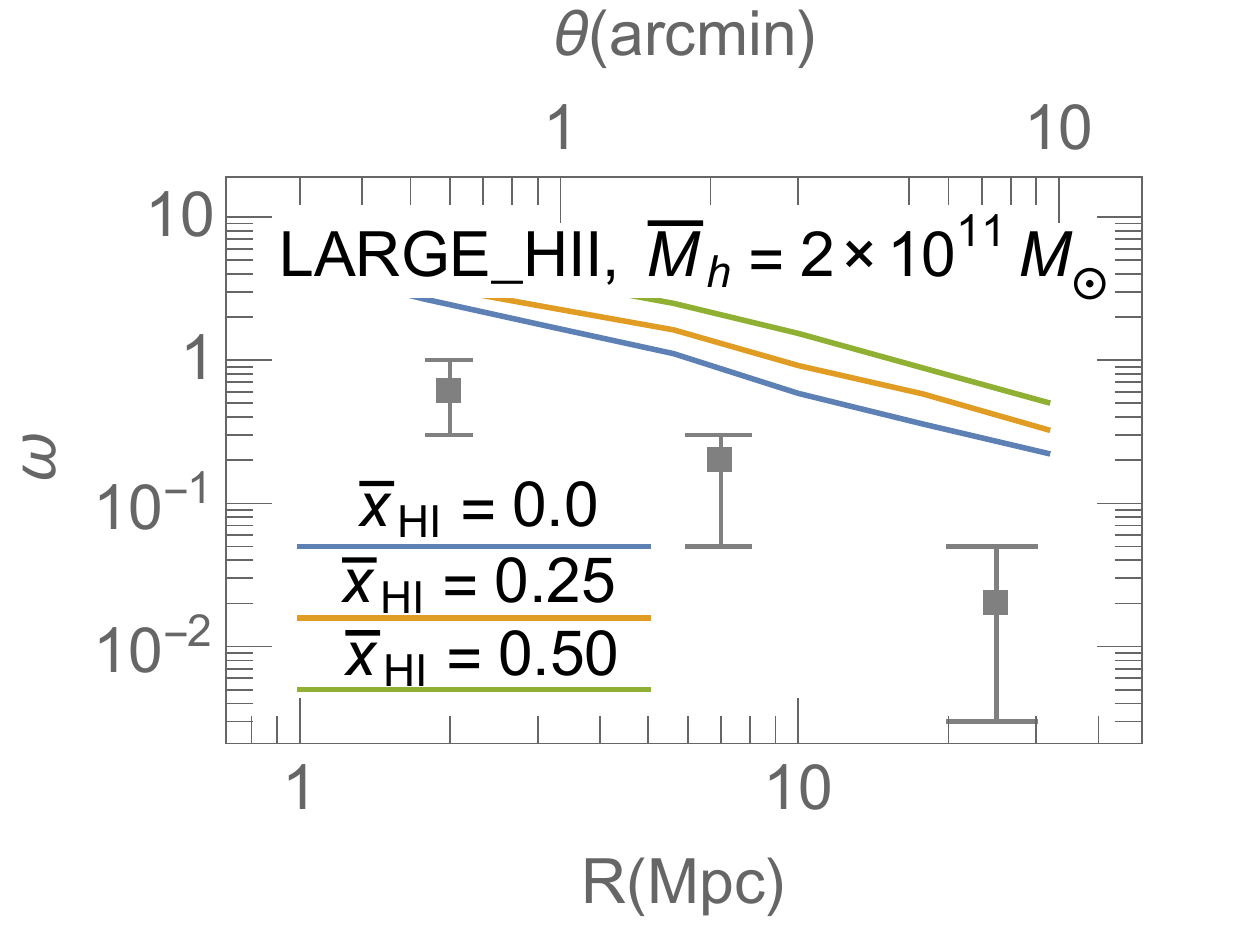} 
\includegraphics[width=0.45\textwidth]{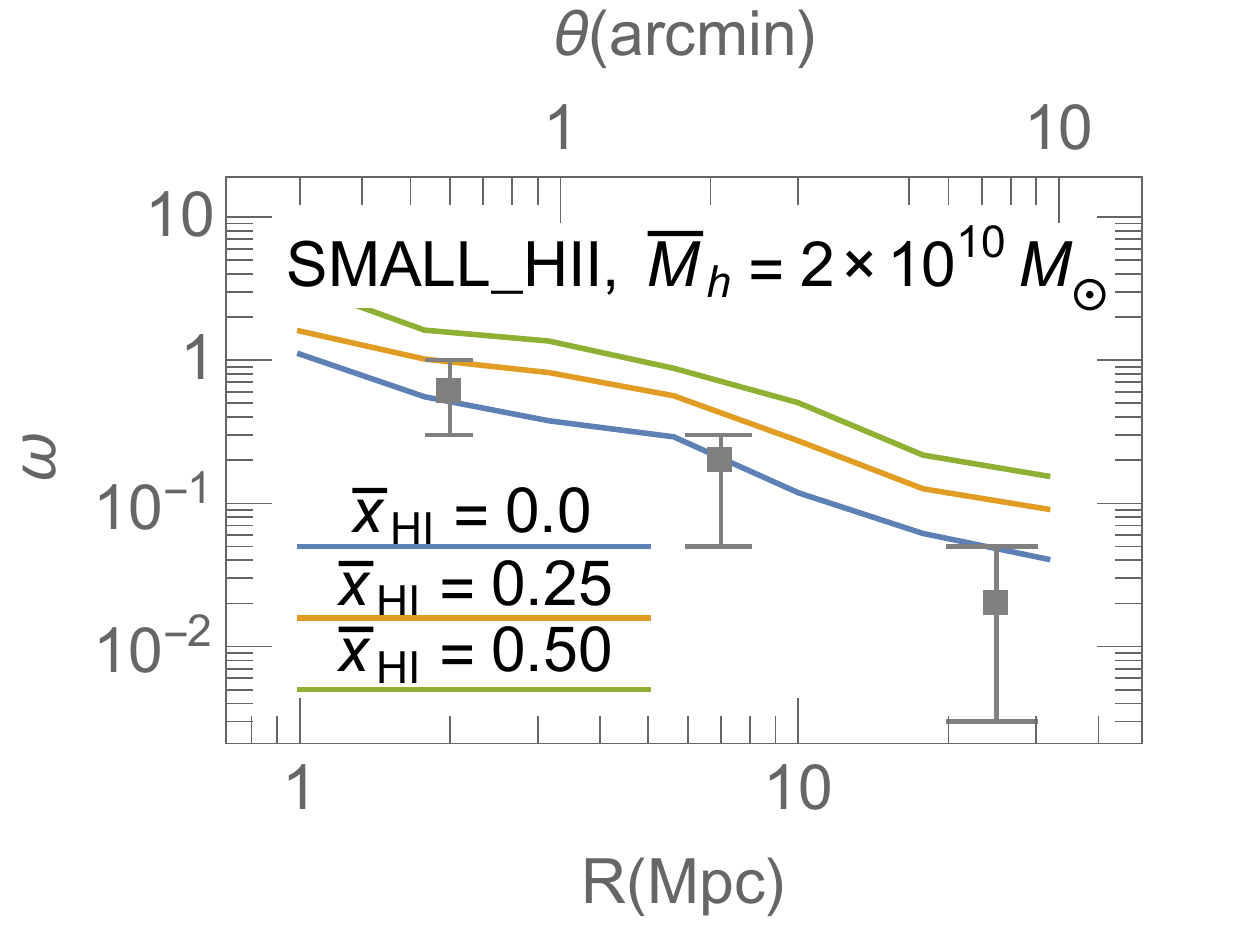}
\includegraphics[width=0.45\textwidth]{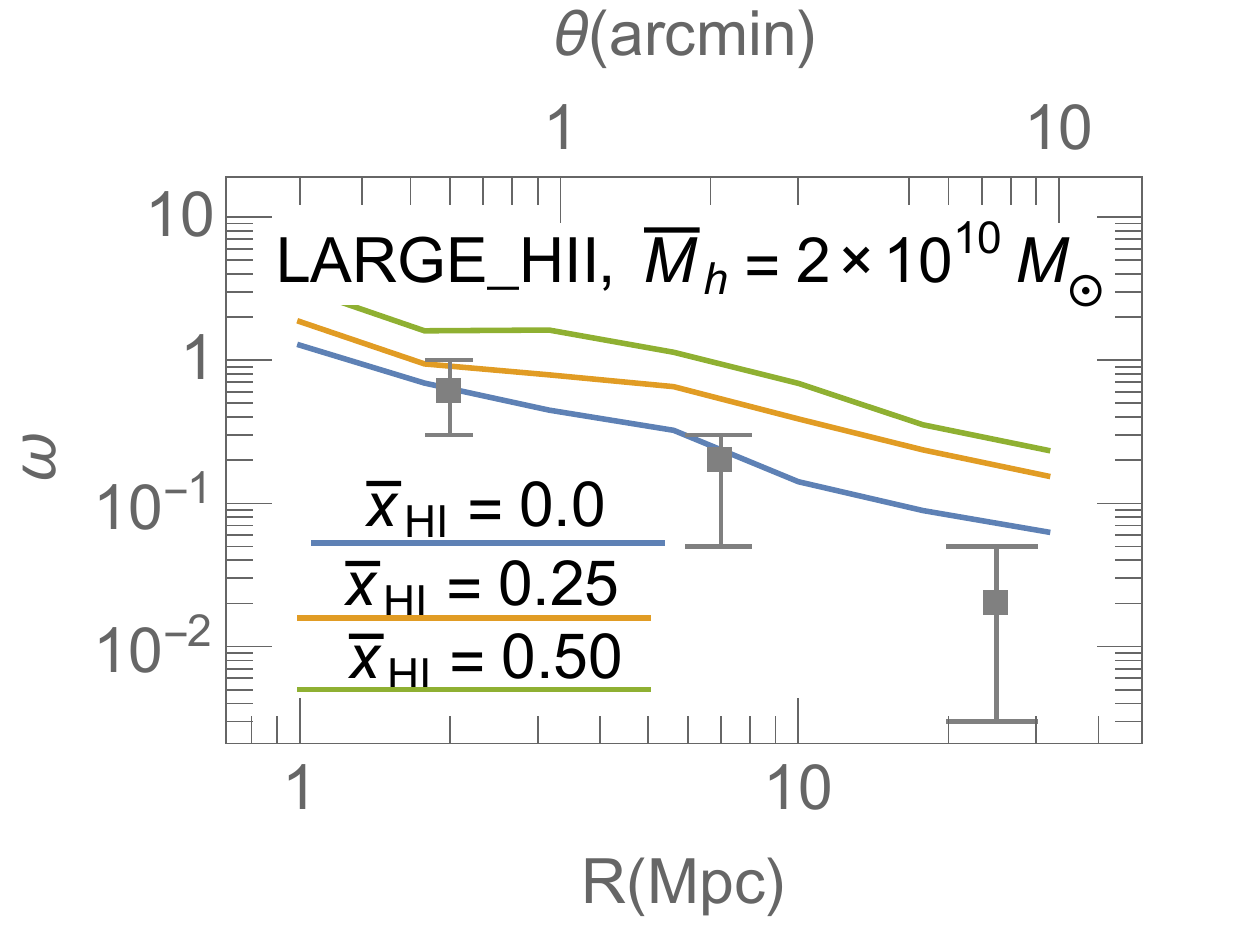} 
\includegraphics[width=0.45\textwidth]{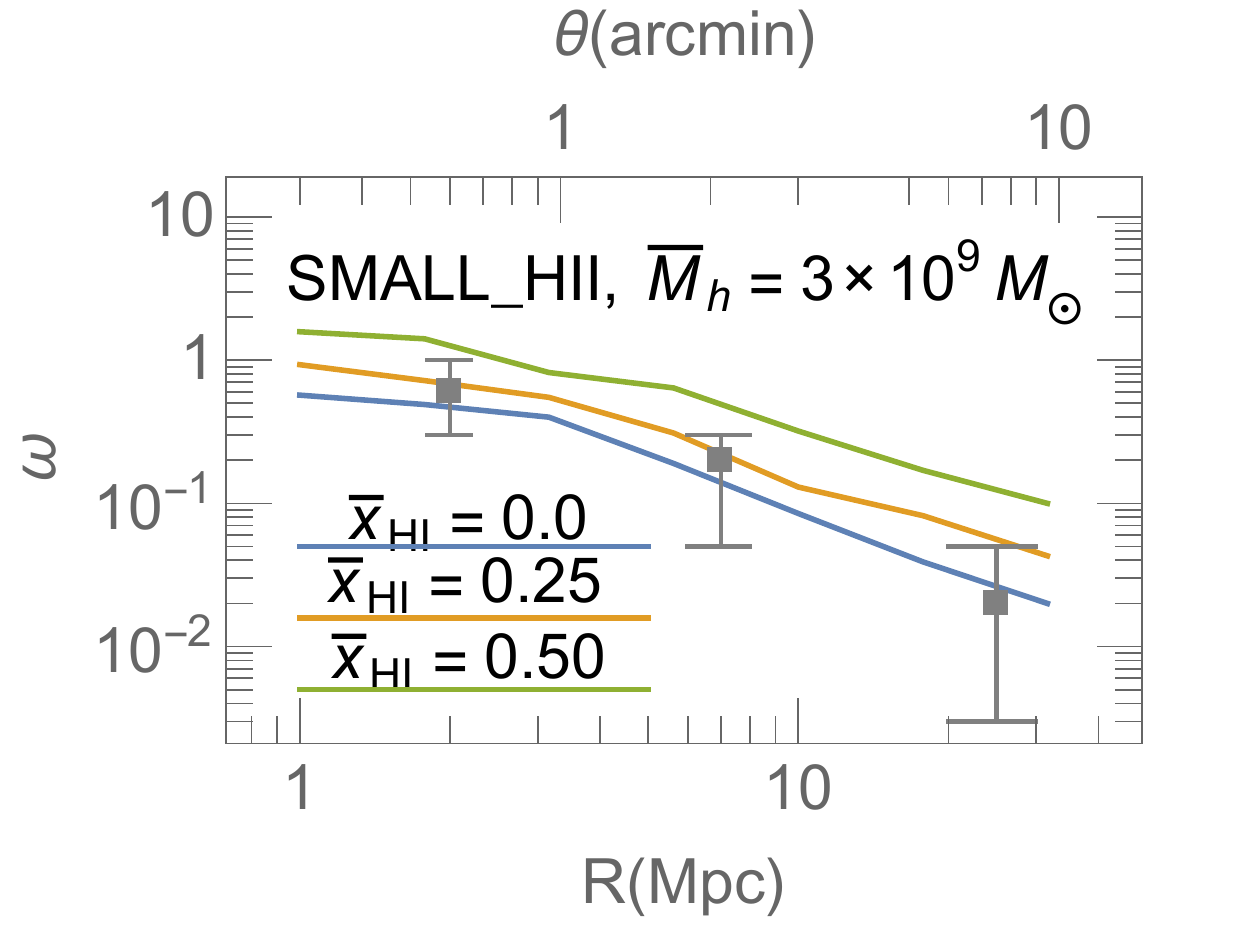}
\includegraphics[width=0.45\textwidth]{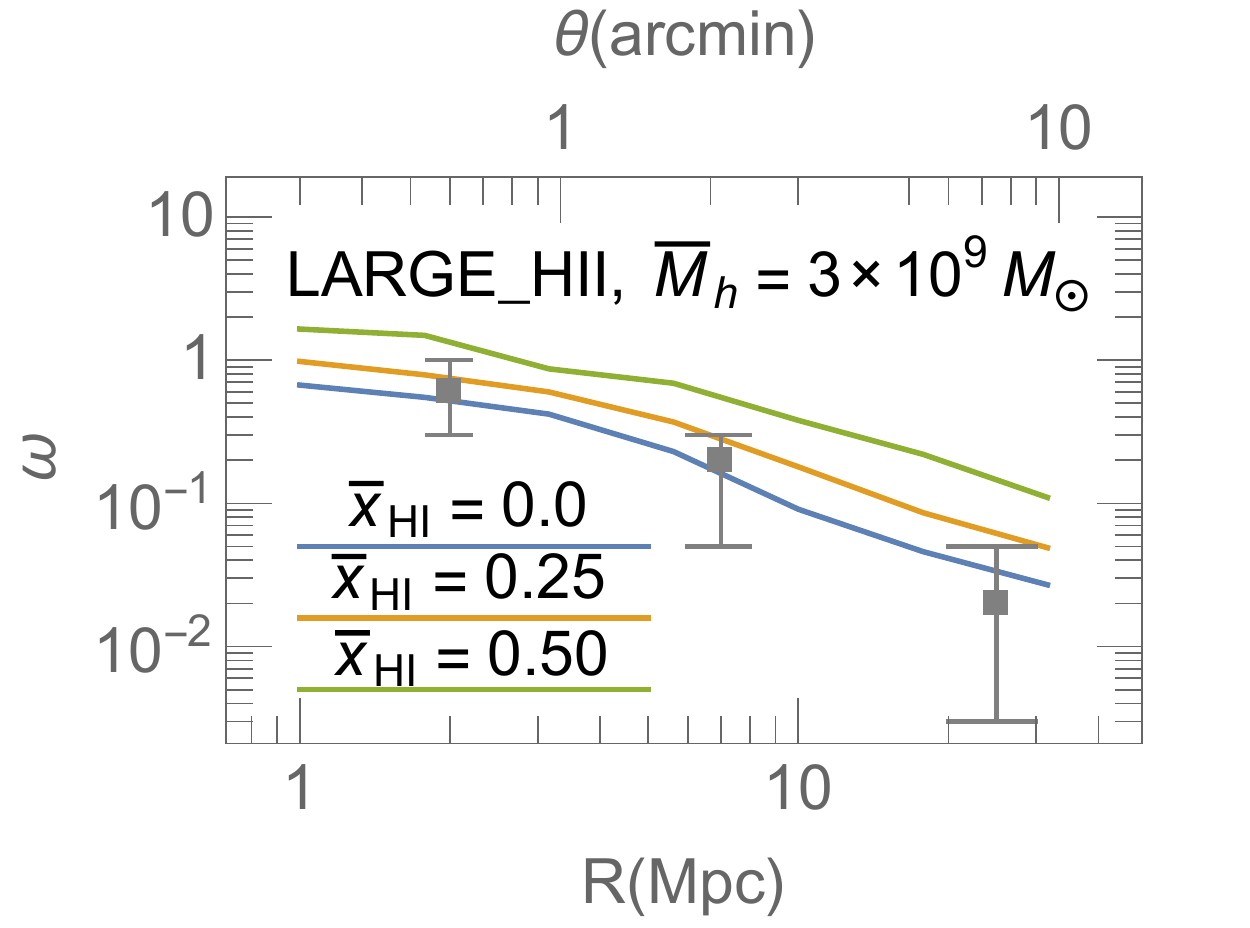} 
}
\caption{Angular correlation functions of the LAEs in out mock survey at $z=6.6$ and $\nlea=4.1\times10^{-4}$ Mpc$^{-3}$, assuming $\bar{x}_{\rm HI}=0$, $\bar{x}_{\rm HI}=0.25$ and $\bar{x}_{\rm HI}=0.5$ ({\it bottom to top in each panel}). The impact of the different reionization morphologies is illustrated by comparing the left and right columns, while in the rows we show the three intrinsic $L_\alpha^{\rm intr} \leftrightarrow M_{\rm h}$ relations.  Points correspond to measurements from Ouchi et al. (2010), with $L_\alpha^{\rm min} = 2.5\times10^{42}$ erg s$^{-1}$.
\label{fig:LAE_comp}
}
\vspace{-1\baselineskip}
\end{figure*}

In Fig. \ref{fig:LAE_comp} we show the angular correlation functions corresponding to our models. The x-axis show the comoving projected separation; for reference $10$ Mpc corresponds to an angular separation of $\sim3$ arcmin.
The impact of the different reionization morphologies is illustrated by comparing the left and right columns, while in the rows we show the three intrinsic $L_\alpha^{\rm intr} \leftrightarrow M_{\rm h}$ relations.
As already noted, the extreme ``Most massive halos'' model with a duty cycle of unity is inconsistent with current observations even post-reionization.  The most massive halos cluster too strongly compared with the observed LAEs.

The other two intrinsic clustering models (bottom two rows) are consistent with the \citet{Ouchi10} observations, for a mostly ionized Universe.  The correlation functions are boosted by a factor of $\sim$ three when the Universe is half ionized (compared with post-reionization).
On the other hand, the morphology of reionization at fixed $\avenf$ (c.f. left and right panels) has a small impact on the clustering of LAEs, with the correlation function  changing at the level of $\sim 10\%$.

It is interesting to notice that a higher $\bar{x}_{\rm HI}$ shifts the correlation function without changing its shape\footnote{The real-space correlation function could show a change in slope if the bubble size distribution is sharply peaked (e.g. \citealt{FZH06}).  However in practice the width of the cosmic HII region PDF combined with the relatively wide redshift bins, $\Delta z\approx 0.1$ of narrow-band LAE surveys smear out such a feature.}  (e.g. \citealt{Jensen13}). Thus the correlation function is robustly described by its value at a given scale; below we make use of this and adopt for our observed statistic the value of the ACF at $10\text{ Mpc}$, $\omega_{10}$.

\begin{figure*}
\vspace{+0\baselineskip}
{
\includegraphics[width=0.7\textwidth]{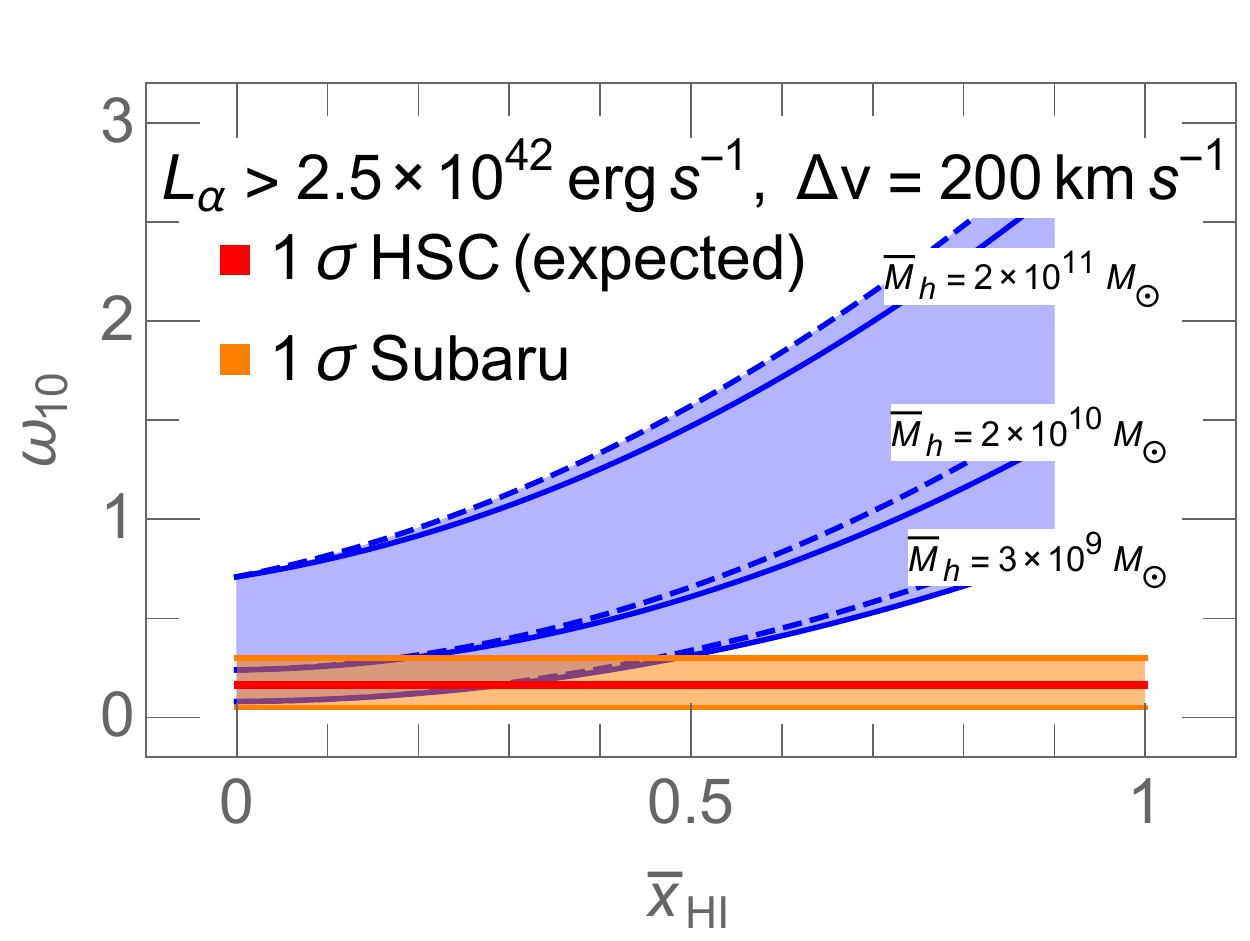}
}
\caption{Relation between the average neutral fraction $\bar{x}_{\rm HI}$ and the angular correlation function at $10\text{ Mpc}$, $\omega_{10}$, with an observed LAE number density of $\nlea = 4.1\times10^{-4}$ Mpc$^{-3}$ and $L_\alpha^{\rm min} = 2.5\times10^{42}$ erg s$^{-1}$ at $z\approx 6.6$. Solid (dashed) blue curves correspond to the  \textbf{SMALL\pmb{\_}HII} (\textbf{LARGE\pmb{\_}HII}) EoR models.  As stated in the text, the IGM absorption is evaluated $\Delta v = 200$ km s$^{-1}$ redward of the systemic redshift. The horizontal orange strip corresponds to the 1-$\sigma$ uncertainty from Ouchi et al. (2010; $L_\alpha^{\rm min} = 2.5\times10^{42}$ erg s$^{-1}$).  The width of the solid red horizontal line corresponds to the expected 1-$\sigma$ uncertainty from the planned Ultra Deep campaign with the HSC.  This figure can be interpreted as follows. The blue shaded region corresponds to the theoretically-allowed range of ACF values; the horizontal strip/line corresponds to the observed/estimated value from Suprime-Cam/Hyper Suprime-Cam.  The region of overlap represents the available constraints from LAE clustering.
\label{fig:LAE_main}}
\vspace{-1\baselineskip}
\end{figure*}

\subsection{Joint constraints on reionization and host halo properties}

In Figure \ref{fig:LAE_main} we show the main result of this work: the relation between the average neutral fraction $\bar{x}_{\rm HI}$ and the angular correlation function at $R=10\text{ Mpc}$, $\omega_{10}$.  Solid (dashed) blue curves correspond to the  \textbf{SMALL\pmb{\_}HII} (\textbf{LARGE\pmb{\_}HII}) EoR models.  The horizontal orange strip corresponds to the 1-$\sigma$ uncertainty from Ouchi et al. (2010; $L_\alpha^{\rm min} = 2.5\times10^{42}$ erg s$^{-1}$).\footnote{These observations assume that the LAE sample is not contaminated by lower redshift interlopers, as argued by \citet{Ouchi10}; however, if the fraction of contaminants is $f_{\rm c}$, the ACF is artificially suppressed by a factor $(1-f_{\rm c})^2$. In Figure \ref{fig:LAE_main_interlop} we show that a reasonable choice for $f_{\rm c}=0.15$ does not affect our conclusions.}  The width of the solid red horizontal line corresponds to the expected 1-$\sigma$ uncertainty from the planned Ultra Deep campaign with the Subaru Hyper Suprime-Cam (HSC), which will probe a comoving volume of $\sim3\times10^6$ Mpc$^3$ at $z\approx6.6$ at the same sensitivity (M. Ouchi, private communication).
The exact position of this line is still to be determined: in the figure, to highlight the improvement in the determination of $x_{\rm HI}$ from clustering measurements by HSC, we have (somewhat arbitrarily) assumed that the mean value of $\omega_{\rm 10}$ does not change, despite a significant decrease in the 1-$\sigma$ uncertainty.
For simplicity and clarity of presentation, we do not include the sample variance of the reionization morphology in observational uncertainties, which depends on the given EoR model \citep{TL14}.  This EoR sample variance should be negligible in the upcoming wide-field HSC survey \citep{Jensen14}.

Figure \ref{fig:LAE_main} can be interpreted as follows.  The blue shaded region corresponds to the theoretically-allowed range of ACF values; the horizontal orange strip corresponds to the observed value.  The region of overlap represents the currently available constraints from LAE clustering.   We can immediately see that even the current observations are very constraining on {\it both} reionization {\it and} the average host halos of LAEs.

Even using our extremely conservative intrinsic clustering model (``Least massive halos'' with $\bar{M}_{\rm h}\approx3\times10^{9}\Msun$) we see that current 1-$\sigma$ constraints from Subaru imply a mostly ionized Universe at $z=6.6$: $\avenf<0.5$ (see also \citealt{McQuinn07LAE} and \citealt{Ouchi10}).   Assuming that the mean value of the ACF remains unchanged, this limit should tighten to $\avenf<0.3$ with the upcoming HSC Ultra Deep Survey.  Note that the less-conservative, intermediate model for the intrinsic clustering, $\bar{M}_{\rm h}\approx2\times10^{10}\Msun$ already implies $\avenf<0.2$.

Clustering measurements also strongly constrain the typical host halos of LAEs.  Even assuming a fully-ionized Universe, the host halo masses of LAEs must be $\bar{M}_{\rm h}\lsim$ few$\times10^{10}\Msun$, in order to be consistent with the Subaru measurements at 1$\sigma$.  The resulting LAE duty cycles, obtained by matching the observed number densities, must be at the percent level or less.  These values can be compared with the analogous ones for $z\approx7$ LBGs, which have $\bar{M}_{\rm h} \sim$ few$\times10^{11}\Msun$ and duty cycles close to unity  (as computed from their two-point correlation functions; \citealt{Barone-Nugent14}).  The disparity between these values motivates the intriguing possibility that $z\approx7$ LAEs and LBGs are dominated by different classes of galaxies.

\subsubsection{Implications for the relation between LAEs and LBGs at $z\approx7$}

Because the HMFs and LAE LFs are steep, the bulk of the clustering signal will be sourced by galaxies close to the detection thresholds.
Therefore the simplest explanation of the different clustering properties of LAEs and LBGs, is that the LAEs with $L_\alpha^{\rm min} \approx 3\times 10^{42}$ erg s$^{-1}$ correspond to UV magnitudes significantly fainter than current Hubble limits of $M_{\rm UV} \approx -18$.  The ratio of \lya\ and UV luminosities is related via the rest frame EW.  Using a probability distribution of EW given a UV magnitude, $P(EW | M_{\rm UV})$, empirically-derived from a large sample of $z\sim3$ LBGs, \citet{Gronke15} estimate that $L_\alpha^{\rm min} \approx 3\times 10^{42}$ erg s$^{-1}$ galaxies should on average have UV magnitudes of  $M_{\rm UV} \approx -18$  at $z\sim6$, making them detectable with current Hubble surveys (see their Fig. 3; see also \citealt{DW12}).  These empirical trends are consistent with theoretically motivated relations between a constant SFR and the corresponding \lya\ and UV luminosities (e.g. \citealt{Kennicutt98,MPD98,  OF06, Garel15}).
  Thus one could expect that the mass estimates of  LAEs and LBGs host halos should be more or less comparable, driven by the clustering properties of galaxies not much brighter than $L_\alpha^{\rm min} \approx 3\times 10^{42}$ erg s$^{-1}$ and $M_{\rm UV} = -18$.
 Instead, the observed $z\approx7$ LBGs have  over an order of magnitude higher average host halo masses and duty cycles than the Subaru LAEs \citep{Ouchi10, Barone-Nugent14}.

One explanation to this apparent discrepancy is if the narrow-band selected LAEs are dominated by a different class of objects than those selected in broad-band dropout surveys.  Although EW distributions have only been constructed from a small, bright sub-sample of Subaru LAEs \citep{Kashikawa11}, they indeed appear to be much flatter than those constructed from color-selected galaxies (e.g. \citealt{DW12, Schenker14}).

  A physical motivation for these trends could be that the narrow-band selected LAEs are dominated by a population of young, star-burst galaxies, residing in less massive halos.  Since the Ly$\alpha$ luminosity of a galaxy is more sensitive to a younger stellar population than its 1500 \AA\ UV luminosity, bursty-galaxies have higher EWs.  Indeed, a bursty SFR seems needed to explain the observed population of high EW galaxies at $z\sim3$ in the models of \citet{Garel15}.   Spectroscopic and photometric UV follow-up of the current and upcoming Subaru fields could help shed light on these trends.

\section{Conclusions}
\label{sec:concl}

Recent observations suggest that both the fraction of drop-out galaxies with strong \lya\ emission and narrow-band selected LAEs show a drop in these populations at $z\gsim6$, reversing empirical trends from lower redshifts. These trends are qualitatively consistent with expectations from the EoR. However, quantifying the corresponding EoR constrains is difficult due to our ignorance of the intrinsic \lya\ emission, and small-number statistics in galaxy samples (e.g. \citealt{Mesinger15} and references therein).

Motivated by the upcoming wide-field survey with the HSC, here we focus on a complementary EoR probe: the clustering of LAEs.  The most important advantage in using the clustering of LAE is that the intrinsic ACFs of dark matter halos are well understood, and only vary by factors of few over the range of possible host halo masses.  This simplifies disentangling the uncertainties from the intrinsic emission and the EoR, allowing us to draw robust conclusions from a single redshift LAE survey.

We confirm (e.g. \citealt{Jensen14}) that the ACF at a fixed observed LAE number density and $\avenf$ is extremely insensitive to the EoR morphology; distinguishing between different EoR models would therefore require more accurate redshift determinations with spectroscopy. We illustrate this possibility with Fig. \ref{fig:LAE_main_narrow} in the Appendix.

Exploring a wide range of possible host halo masses, we conclude that current measurements of the $z=6.6$ ACFs \citep{Ouchi10} imply $\avenf \lsim 0.5$ (1-$\sigma$).  Although in qualitative tension with the \lya\ fraction measurements which favor a neutral Universe, these two complementary probes are consistent at 1-$\sigma$ (e.g. \citealt{Caruana14, Dijkstra14, Mesinger15}).  The upcoming Ultra Deep campaign with the HCS could improve on these constraints by tens of percent, constraining $\avenf\lsim0.3$ (1-$\sigma$) if the mean value of the ACF remains unchanged.

Since the value of the LAE ACF is relatively low, it is possible to put an upper limit on the masses of their host halos (the EoR can only increase the value of the observed ACFs). We find that $\bar{M}_{\rm h} \lsim 10^{10} \Msun$, an order of magnitude lower than the analogous value for color-selected, LBGs at $z\approx7$. Combined with their observed number densities, this implies a very low duty cycle $\lsim$ few per cent. We suggest that this discrepancy could be due to the LAEs selection technique, preferentially selecting a population of young, star-burst galaxies, residing in less massive halos.

\vskip+0.3in

We thank M. Dijkstra and M. Trenti for useful comments on an earlier version of this manuscript.  We are extremely grateful to M. Ouchi for providing us estimates for the upcoming HCS survey.

\bibliographystyle{mn2e}
\bibliography{ms}

\appendix
\section{Sensitivity of the Angular Correlation Functions to model parameters}
\label{sec:appendix}

\begin{figure}
\vspace{+0\baselineskip}
{
  \includegraphics[width=0.45\textwidth]{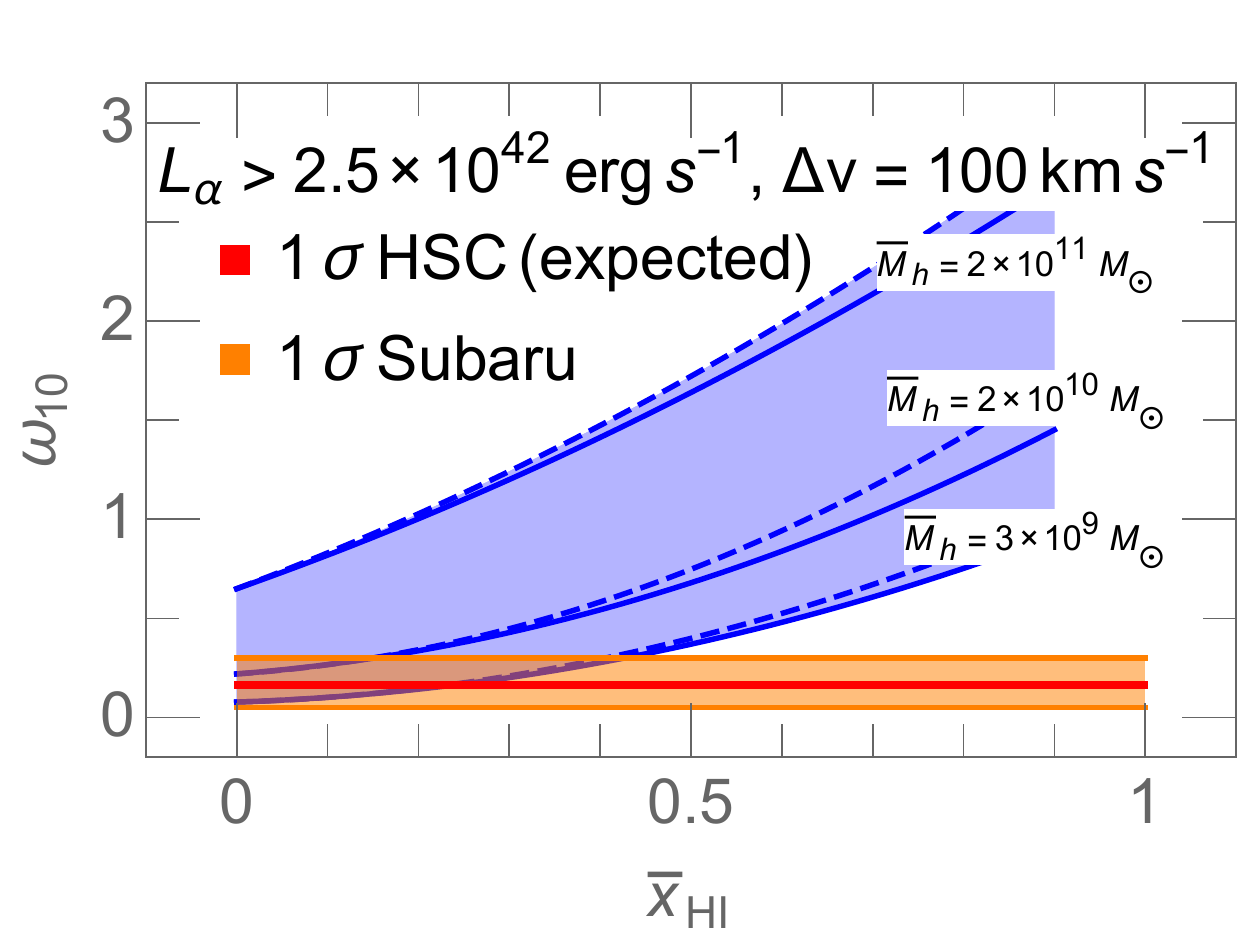}
}
\caption{Same as Fig. \ref{fig:LAE_main}, but assuming a systemic velocity offset of the \lya\ line of $\Delta v=100\text{ km s$^{-1}$}$. The similarity of this figure and Fig. \ref{fig:LAE_main} demonstrates that our results are not strongly affected by the poorly-understood \lya\ line profiles.
\label{fig:LAE_main_100}}
\vspace{-1\baselineskip}
\end{figure}

\begin{figure}
\vspace{+0\baselineskip}
{
  \includegraphics[width=0.45\textwidth]{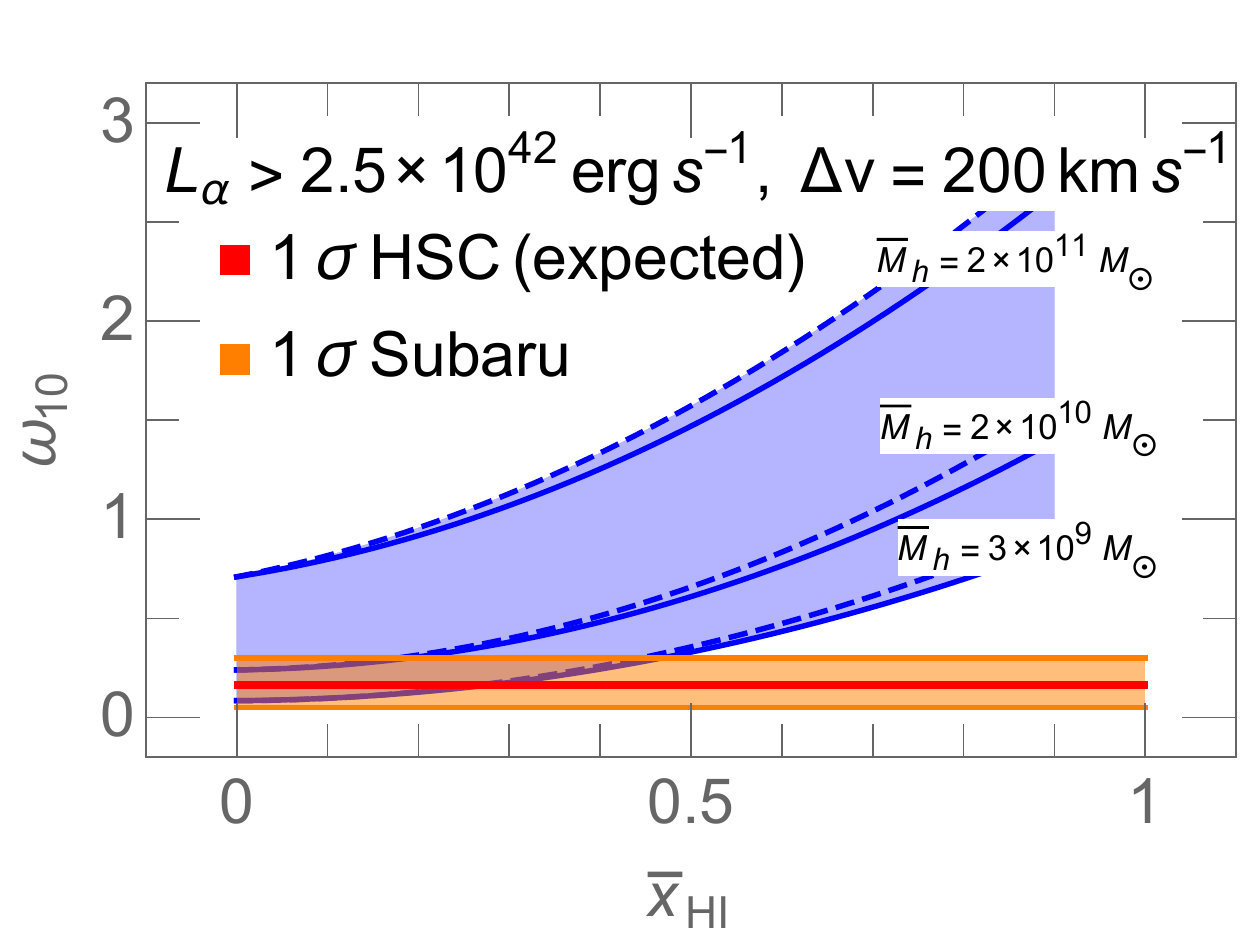}
}
\caption{Same as Fig. \ref{fig:LAE_main}, but assuming 
a $L_{\alpha}^{\rm intr}\propto M_{\rm h}^{2/3}$ scaling for the ``Least massive halos'' curve. The similarity of this figure and Fig. \ref{fig:LAE_main} demonstrates that our results at fixed $\bar{M}_{\rm h}$ are not affected by the precise mass scaling of the $L_{\alpha}^{\rm intr} \leftrightarrow M_{\rm h}$ relation, since the clustering signature is dominated by halos with masses close to $\bar{M}_{\rm h}$.  We do note that these halo mass--to--luminosity scalings, if extrapolated to the bright end of the LFs, would result in very different predictions (see Fig. \ref{fig:LAE}).
\label{fig:LAE_main_lowbeta}}
\vspace{-1\baselineskip}
\end{figure}

\begin{figure}
\vspace{+0\baselineskip}
{
\includegraphics[width=0.45\textwidth]{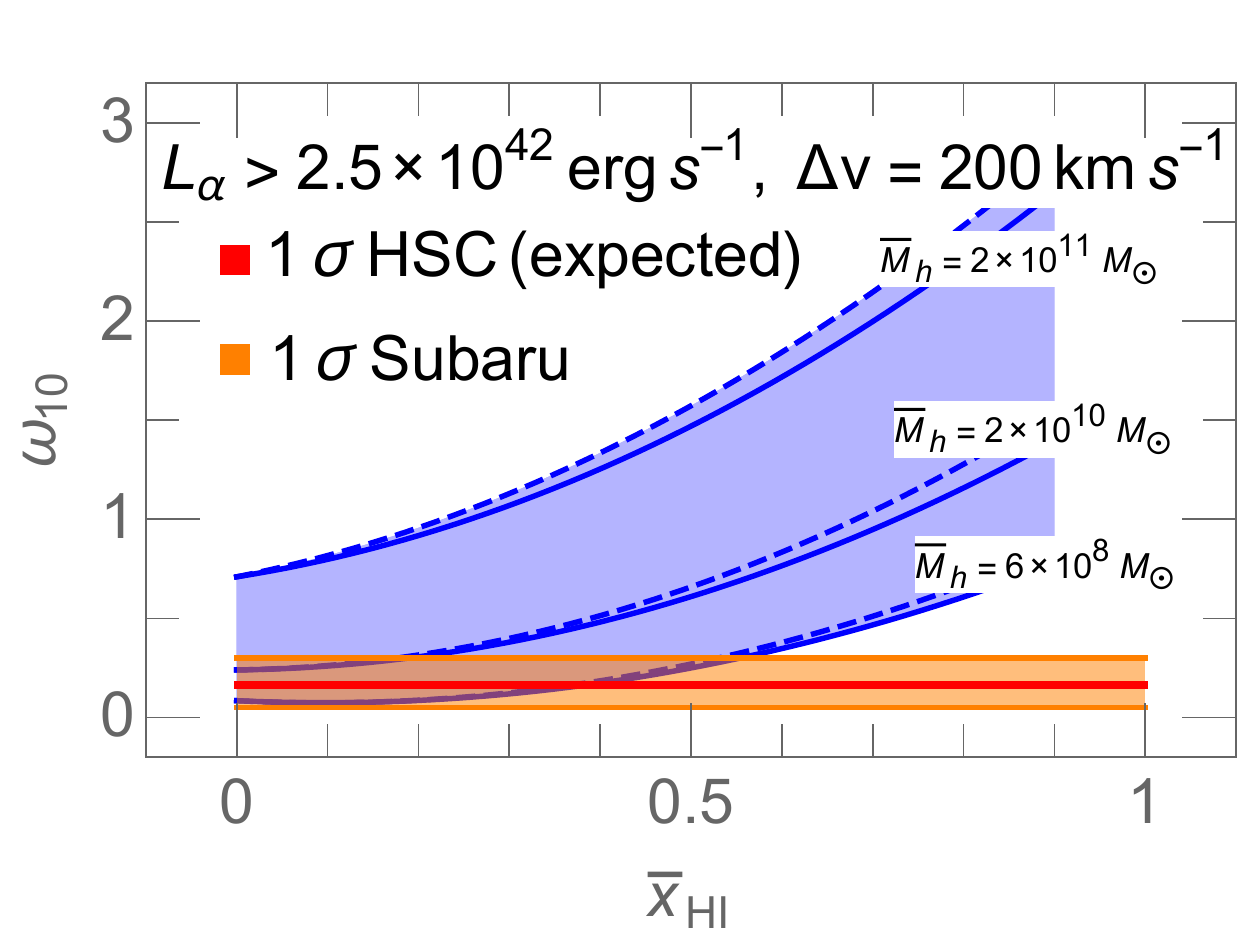}
}
\caption{Same as Fig. \ref{fig:LAE_main}, but $\bar{M}_{\rm h}=6\times 10^8 M_\odot$ for the ``Least massive halos'' curve. The similarity of this figure and Fig. \ref{fig:LAE_main} demonstrates that our results are marginally affected by a further decrease of $\bar{M}_{\rm h}$ below $3\times 10^9 M_\odot$.
\label{fig:LAE_main_lowM}}
\vspace{-1\baselineskip}
\end{figure}

\begin{figure}
\vspace{+0\baselineskip}
{
\includegraphics[width=0.45\textwidth]{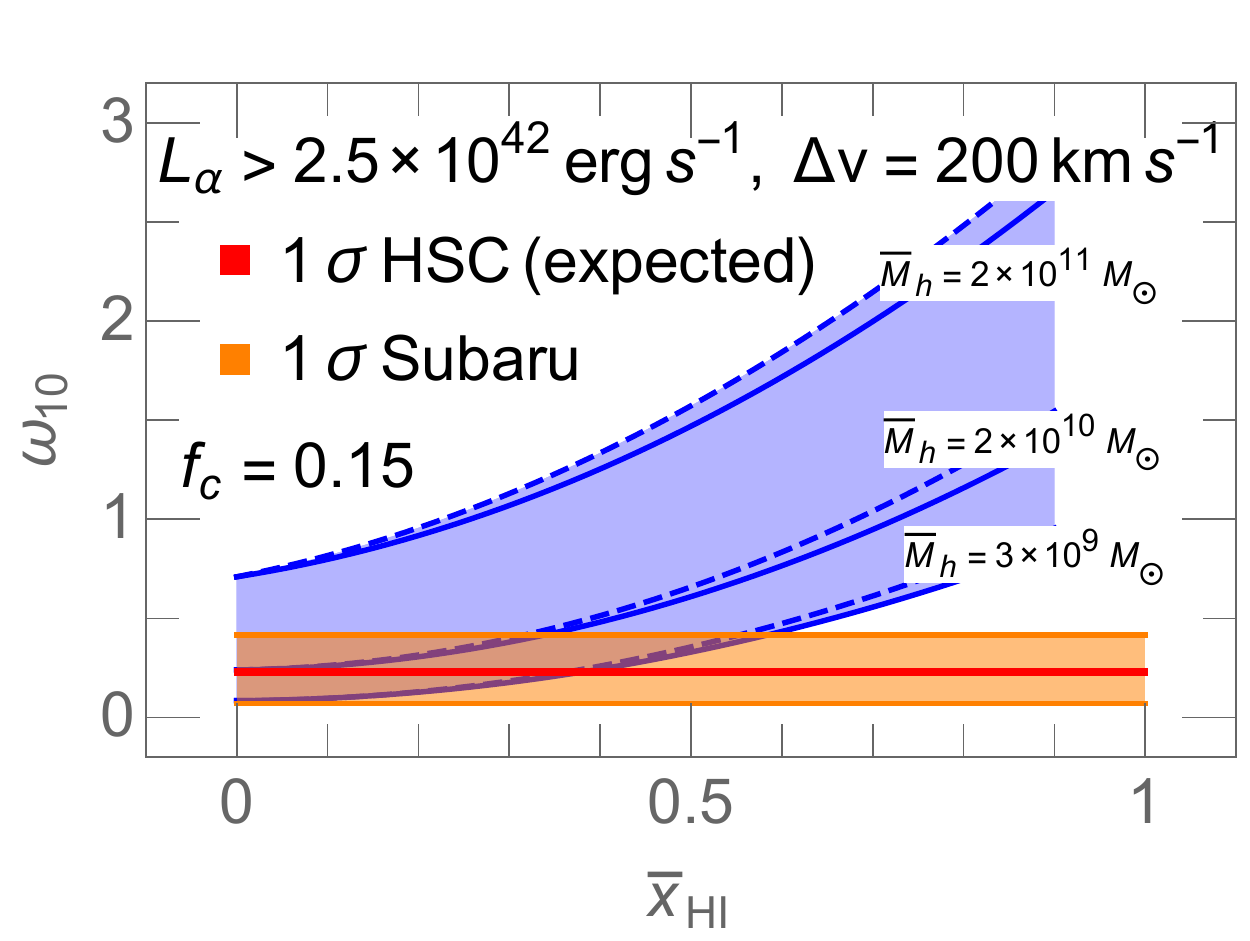}
}
\caption{Same as Fig. \ref{fig:LAE_main}, but assuming a fraction $f_{\rm c}=0.15$ of randomly-distributed, lower redshift interlopers when plotting the horizontal strip corresponding to Subaru observations. Although the allowed range of $\bar{M}_{\rm h}$ values broadens somewhat, the main conclusions remain unchanged.
\label{fig:LAE_main_interlop}}
\vspace{-1\baselineskip}
\end{figure}

\begin{figure}
\vspace{+0\baselineskip}
{
\includegraphics[width=0.45\textwidth]{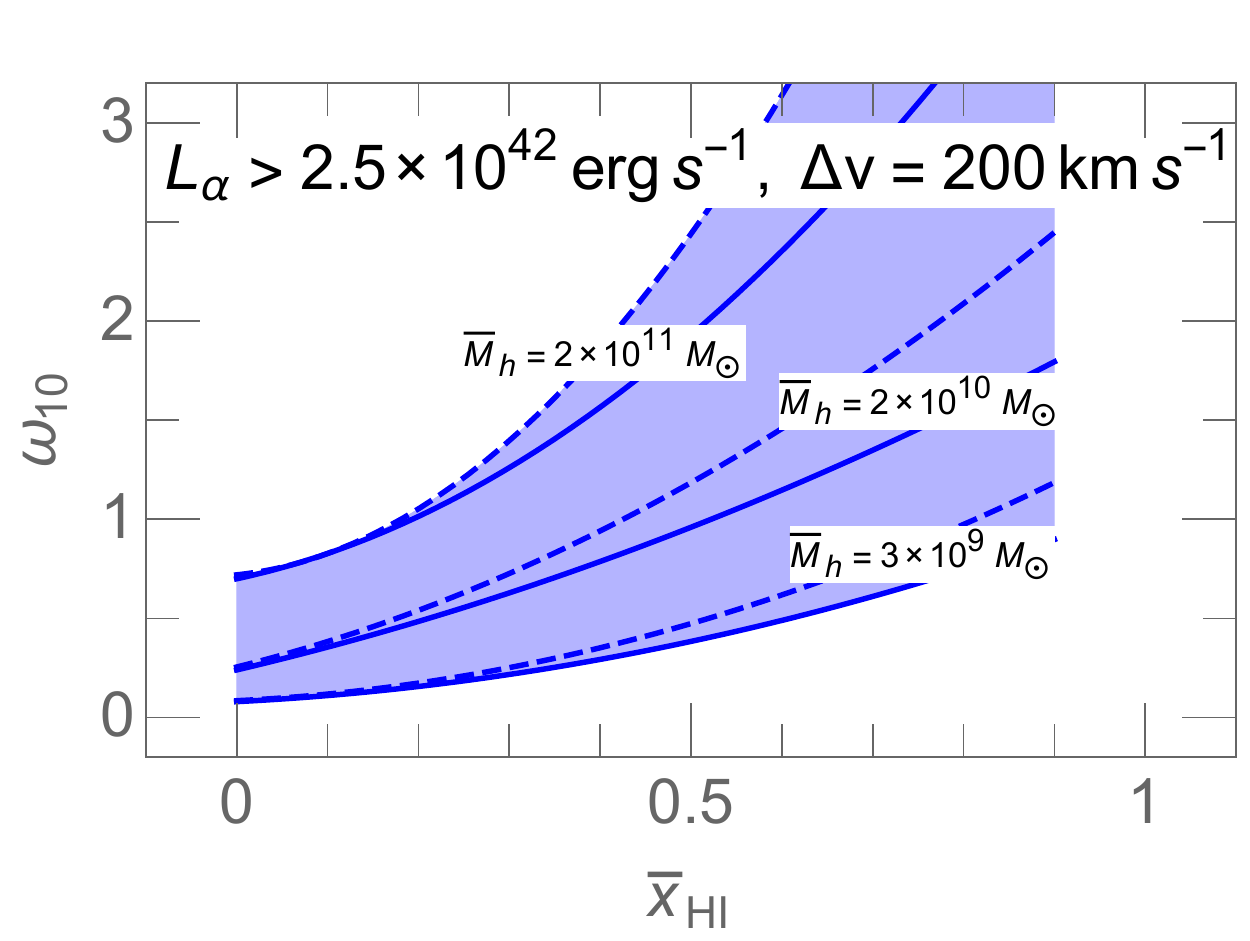}
}
\caption{Predictions for $\omega_{10}$ as in Fig. \ref{fig:LAE_main}, but assuming a survey depth of $\Delta z = 0.014$ (instead of $\Delta z = 0.1$), as could be achieved with \lya\ spectroscopy.  The dependence of $\omega_{10}$ on the EoR morphology is greater.
\label{fig:LAE_main_narrow}}
\vspace{-1\baselineskip}
\end{figure}

In this appendix, we show some alternate versions of Fig. \ref{fig:LAE_main}, varying other fiducial parameter choices.  In Fig. \ref{fig:LAE_main_100} we vary the velocity offset from the systemic redshift assuming $\Delta v=100\text{ km s$^{-1}$}$, in Fig. \ref{fig:LAE_main_lowbeta} we assume a $L_{\alpha}^{\rm intr}\propto M_{\rm h}^{2/3}$ scaling for the ``Least massive halos'' curve, and in Fig. \ref{fig:LAE_main_lowM} we use $\bar{M}_{\rm h}=6\times 10^8 M_\odot$ for the ``Least massive halos'' curve. The similarity of these figures and Fig. \ref{fig:LAE_main} demonstrates that our conclusions are not notably affected by these parameters.

In Fig. \ref{fig:LAE_main_interlop} we instead change the horizontal strip in order to show the impact of $f_{\rm c}$ = 15\% low-redshift interlopers in the Subaru field.  The maximum impact of interlopers (assuming a random distribution) would be an artificial suppression of the ACF by a factor $(1-f_{\rm c})^2$. Although \citet{Ouchi10} argue for  $f_{\rm c} \sim$ 0\%, and find no evidence of contaminants in ongoing spectroscopic follow-up, they quote a strict upper limit of $f_{\rm c}\le 30\%$.  In the case of interloper contamination, the allowed range of $\bar{M}_{\rm h}$ values broadens somewhat, but our main conclusions remain unchanged.

Finally, in Fig. \ref{fig:LAE_main_narrow} we show how LAEs can probe reionization morphology, provided they are better localized in redshift space.  In this figure, we take a survey depth of 5 Mpc ($\Delta z = 0.014$ instead of $\Delta z = 0.1$).  The curves for the two EoR models diverge much more than in the previous figures.  This survey depth (corresponding to $\Delta v \sim 600$ km s$^{-1}$ in the rest frame) is broader than the observed spread in the systemic line offsets (e.g.  \citealt{Shibuya14, Stark15, Sobral15}); hence such measurements can be achieved without metal line detections.  These results illustrate that wide-field LAE spectroscopy, either through follow-up or through dedicated instruments such as the Subaru Prime Focus Spectrograph, could be very powerful in discriminating between EoR models.

\end{document}

%% file: defns.tex
\newcommand{\Mhalo}{M_h}
\newcommand{\nlea}{n_{\rm LAE}}

\newcommand{\cmfastSM}{\textsc{\small 21CMFASTv2}}
\newcommand{\cmfast}{\textsc{\small 21CMFAST}}

\newcommand{\avenf}{\bar{x}_{\rm HI}}

\newcommand{\lya}{Ly$\alpha$}

\newcommand{\Msun}{M_\odot}

\newcommand\lsim{\mathrel{\rlap{\lower4pt\hbox{\hskip1pt$\sim$}}
        \raise1pt\hbox{$<$}}}
\newcommand\gsim{\mathrel{\rlap{\lower4pt\hbox{\hskip1pt$\sim$}}
        \raise1pt\hbox{$>$}}}
\def\myputfigure#1#2#3#4#5%
{\vskip#5pt\makebox[0pt]{\hskip#2in
\includegraphics[width=#3\textwidth]{#1}}\vskip#4pt\hfill}

